\documentclass[a4paper,12pt]{article}
\pdfoutput=1
\usepackage{amsfonts,amssymb,amsmath,amssymb}
\usepackage{graphics,graphicx}
 
\newcommand{\ba}{\begin{array}}
\newcommand{\ea}{\end{array}}
\newcommand{\beq}{\begin{equation}}
\newcommand{\eeq}{\end{equation}}

\def\pbn3{\mathcal S_{bn}}
\def\pgn3{\mathcal S_{gn}}

\def\f2r{\mathrm{Ref_{2R}}}
\def\rr{\mathrm{Re\rho_{2}}}
\def\ri{\mathrm{Im\rho_{3}}}
\def\onebmw2{\frac{1}{m_W^2}}
\def\mt{m_t}

\def\shat{\hat{s}}
\def\that{\hat{t}}
\def\uhat{\hat{u}}

\def\bt{\begin{table}}
\def\et{\end{table}}
\def\bc{\begin{center}}
\def\ec{\end{center}}
\def\bi{\begin{itemize}}
\def\ei{\end{itemize}}
\def\bea{\begin{eqnarray}}
\def\eea{\end{eqnarray}}
\def\beas{\begin{eqnarray*}}
\def\eeas{\end{eqnarray*}}

\def\fla{\mathrm{f_{1L}}}

\begin{document} 
\begin{flushright} 
 ADP-15-24/T926
\end{flushright} 

\begin{center}
{\Large  \boldmath \bf
Polarization of top quark as a probe of its chromomagnetic and chromoelectric 
couplings in \\\vskip 8pt $tW$ production  at the Large Hadron Collider} 
\\ \vspace*{0.2in} 
{\large Saurabh D. Rindani$^1$, Pankaj Sharma$^2$ and Anthony W. Thomas$^2$} \\
\vspace*{0.2in}
{\small \it
\small \it $^1$ Theoretical Physics Division, Physical Research Laboratory,\\
\small \it Navrangpura, Ahmedabad 380 009, India\\
\small \it $^2$ Center of Excellence in Particle Physics (CoEPP),\\
\small \it University of Adelaide, Adelaide, Australia }
\end{center}
\vspace*{0.6in}
\begin{center}
{\large\bf Abstract}
\end{center}

We study the sensitivity of the Large Hadron Collider (LHC) to top quark chromomagnetic (CMDM) 
and chromoelectric (CEDM) dipole moments and $Wtb$ effective couplings in single-top production in
association with a $W^-$ boson, followed by semileptonic decay of the top. The $Wt$
single-top production mode helps to isolate the anomalous $ttg$ and $Wtb$ couplings, in contrast
to top-pair production and other single-top production modes, where other new-physics effects can
also contribute. We calculate the top
polarization and the effects of these anomalous couplings on it at two centre-of-mass (cm)
energies, 8 TeV and 14 TeV. As a measure of top polarization, we look at decay-lepton angular
distributions in the laboratory frame,  without requiring reconstruction of the rest frame of the
top, and study the effect of the anomalous couplings on these distributions. We construct certain
asymmetries to study the sensitivity of these distributions to top-quark couplings. 
We determine individual limits on the dominant couplings, viz., the real part of
the CMDM $\rr$, the imaginary part of the CEDM $\ri$, and the real part of the tensor $Wtb$
coupling $\f2r$, which may be obtained by utilizing these asymmetries at the LHC. We also obtain
simultaneous limits on pairs of these couplings taking two couplings to be non-zero at a time.

\newpage
\section{Introduction}

The top quark is the heaviest fundamental particle discovered so far with mass $m_t=173.2\pm 0.9$
GeV \cite{top:mass}. Mainly for this reason, it is considered to be one of the most likely places
where new physics might be discovered. While enough information about the top quark is already
available, showing consistency with SM expectations, future runs at the Large Hadron Collider
(LHC) will enable a more precise determination of its properties. In particular, if new physics
contributions to the interactions of the top quark are written in terms of anomalous couplings, it
will be possible to constrain these couplings quantitatively. The experimental study involves the
measurement of production cross sections, kinematic distributions in production and decay
characteristics. With the availability of high statistics, it should also be possible to
investigate finer details like top polarization, which can give more information about its
interactions. 

Because of its large mass, the top-quark life time is very short and it decays spontaneously
before any non-perturbative QCD effects can force it into a bound state. Thus, its spin
information is preserved in terms of the differential distribution of its decay products. So by
studying the kinematical distributions of top decay products, it is, in principle, possible to
measure the top polarization in any top production process. 

Top polarization and its usefulness in the study of new physics scenarios has been extensively 
treated in the literature. For reviews see \cite{Bernreuther:2008ju,Bernreuther:2015wqa}.
Some recent papers in the context of hadron colliders are \cite{Bernreuther:2013aga,
Godbole:2010kr,Huitu:2010ad,Godbole:2011vw,Rindani:2011pk,Prasath:2014mfa,Rindani:2011gt,
Biswal:2012dr,Rindani:2013mqa,top:fba}. For example, in Ref. 
\cite{Godbole:2010kr}, it was shown in the context of an extra $Z$ model how decay-lepton
asymmetries in the lab. frame could be used to measure top polarization. Ref.
\cite{Huitu:2010ad,Godbole:2011vw} studied the top polarization in associated single-top
production with charged Higgs in two Higgs doublet model (2HDM) and the minimal supersymmetric
extension of standard model (MSSM).
The effect of anomalous $Wtb$ couplings on top polarization in single-top production in
association with $W$ has been studied, without \cite{Rindani:2011pk} and  with CP violation
\cite{Rindani:2011gt}. Constraining top-quark chromomagnetic (CMDM) and chromoelectric (CEDM)
dipole couplings using top polarization observables in the context of Tevatron and LHC has been
studied in \cite{Bernreuther:2013aga,Biswal:2012dr}. Ref. \cite{Rindani:2013mqa} discusses the use
of top polarization to determine the charged-Higgs mass and to distinguish various 2HDMs in
associated $tH^-$ production at the LHC. Refs. \cite{top:fba} suggest utilizing top polarization
as a probe of models for the top forward-backward asymmetry observed at the Tevatron.

At the LHC, top quarks are produced mainly via two independent mechanisms. The dominant one is 
$t\bar t$ pair production which occurs through gluon fusion and quark-antiquark annihilation. 
The second mechanism is single top production \cite{Heinson:1996zm,Stelzer:1998ni,
Belyaev:1998dn,Boos:1999dd,Tait:1999cf,Espriu:2001vj,Espriu:2002wx,Tait:2000sh,
White:2009yt,Frixione:2008yi,Frixione:2005vw, Chatrchyan:2011vp,Khachatryan:2014iya,
Aad:2014fwa,Aad:2012xca,Chatrchyan:2014tua}.
Since the latter proceeds via weak interaction, top quarks tend to have large polarization
\cite{Espriu:2002wx,Tait:2000sh}. At LHC energies, single-top quark events in the SM are produced
in three different modes:  a) the $t$-channel ($bq\rightarrow tq^\prime$), b) the $s$-channel 
($q\bar q^\prime\rightarrow t\bar{b}$) and c) the $Wt$ production process 
($bg\rightarrow tW^-$) \cite{Tait:1999cf}. These three modes are completely different
kinematically and can be separated from one another. Of these, the $t$-channel
\cite{Chatrchyan:2011vp,Khachatryan:2014iya,Aad:2014fwa} and the $Wt$ 
\cite{Aad:2012xca,Chatrchyan:2014tua} processes have been observed at the LHC. Despite having
smaller cross section than top pair production, single-top production is an important tool to
study effects which may not be accessible in top pair production. For example, single-top
production allows an independent measurement of the CKM matrix element $V_{tb}$. Also, unlike 
top-pair production, single-top production gives rise to large top-quark polarization. Top
polarization has indeed been measured at the LHC \cite{cms-pas-top-13-001} in the $t$-channel
process, and found to be consistent with the SM prediction, within somewhat large errors.

Run 2 has recently commenced with center of mass (cm) energy of 13 TeV. The LHC is expected to run at 
this energy for about a year collecting around 30 fb$^{-1}$ of integrated luminosity. The next 
run of the LHC will be a high-luminosity run with a slightly increased cm energy of 14 TeV. This run is 
expected to acquire a tremendous amount of data (around 3000 fb$^{-1}$). In this work, 
we focus on the 8 and 14 TeV runs of the LHC and present all the results for them. 
The priorities at the LHC in the analysis of these large data sets would be to first determine 
accurately the total and differential cross sections for the
dominant top-pair production process, followed by those for single-top processes. This will help
to constrain anomalous couplings contributing to the cross sections for these processes. However,
there are competing contributions from several sources. In such a scenario, closer detailed
examination of top polarization and the consequent decay distributions would be helpful in
isolating those sources.

In this work, we study $Wt$ production at the LHC in the presence of 
anomalous gluon couplings to top quarks. In particular, we examine the possibility of using 
top polarization and other angular observables, constructed from top decay products in the
laboratory frame, to measure these couplings. Our main emphasis will be to show how
these laboratory-frame observables can be used to probe the anomalous couplings. 
We calculate our observables at the LHC with  centre-of-mass (cm) energies of 8 TeV (LHC8) 
and 14 TeV (LHC14). We also study the sensitivities of these observables to anomalous 
top quark couplings including statistical uncertainties with integrated luminosities 20 fb$^{-1}$
at LHC8 and 30 fb$^{-1}$ for the case of LHC14. Our results will go through with little change for the present run of the LHC at 13 TeV for a similar luminosity.

Top quark couplings to a gluon can be defined in a general way as
\beq
\Gamma^\mu=\rho_1\gamma^\mu +\frac{2 i}{m_t}\sigma^{\mu\nu}\left(\rho_2+i\rho_3\gamma_5\right)
q_\nu,
\eeq
where $\rho_2$ and $\rho_3$ are top quark CMDM and CEDM form factors and $m_t$ is the mass
of top quark. Of these, the $\rho_2$ term is CP even, whereas the $\rho_3$ term is CP odd. In the
SM, both $\rho_2$ and $\rho_3$ are zero at tree level. Top CMDM and CEDM couplings, which we study
here, could arise in the SM or from new interactions at loop level. While the CP-conserving CMDM
coupling can arise in the SM at one-loop \cite{Martinez:2007qf}, the CP-violating CEDM coupling
can only be generated at 3-loop level in the SM \cite{3-loop}. These couplings have been
calculated at one-loop
level in various new physics models such as MSSM \cite{mssm}, 2HDM \cite{2hdm,Gaitan:2015aia}, 
Little Higgs model \cite{LH} and in models with unparticles \cite{unparticle}.

Top chromomagnetic and chromoelectric dipole moments have been examined in the past in the context of
single-top production \cite{Rizzo:1995uv,Ayazi:2013cba,Fabbrichesi:2014wva}, top-pair production
\cite{Biswal:2012dr,atwood,haberl,hioki,saha,hioki2,hesari,gupta,Fabbrichesi:2013bca}, and 
top-pair plus jet production \cite{Cheung:1995nt} at hadron colliders. Cheung \cite{Cheung:1996kc}
has used spin correlations in top-pair production at hadron colliders to probe top CMDM and CEDM.
CP violation in top-pair production at hadron colliders including top CEDM couplings is studied in
\cite{Zhou:1998wz}.

Apart from having a direct effect on top-pair production at hadron colliders, top CMDM and CEDM
couplings can have an indirect effect and modify the decay rate of $b\to s\gamma$ at loop level
\cite{Hewett:1993em,b2sg}. Using the measured branching ratio Br($b\to s\gamma$) \cite{b2sg},
tight bounds on the top CMDM, $\rho_2$, were extracted, viz.,  $0.03 < \rho_2 < 0.01$.

Any new physics in which new couplings to top are chiral can influence top polarization. The
measurement of top polarization is thus an important tool to study new physics in single-top
production. However, top polarization can only be measured through the distributions of its
decay products. Hence, any new physics in top decay may contaminate the measurement of 
top polarization, and therefore of the new physics contribution in top production. Assuming
only SM particles, any new physics in top decay can be parametrized in terms of anomalous
$tbW$ couplings as
\begin{equation}
 \Gamma^\mu =\frac{-ig}{\sqrt{2}}V_{tb}\left[\gamma^{\mu}(\mathrm {f_{1L} P_{L}}
 +\mathrm {f_{1R} P_{R}})+
\frac{i \sigma^{\mu \nu}}{m_W}(p_t -p_b)_{\nu}(\mathrm {f_{2L} P_{L}}
+\mathrm {f_{2R} P_{R}})\right] \label{anomaloustbW}
\end{equation}
where the in SM $\mathrm {f_{1L}}=1$ and $\mathrm {f_{1R}}=\mathrm {f_{2L}}=\mathrm {f_{2R}}=0$.
Under the assumptions that (i) anomalous $tbW$  couplings are small, (ii) the top is on-shell and
(iii) $t\rightarrow bW^+$ is the only decay channel, it was shown in Refs. \cite{Godbole:2006tq}
that the charged-lepton angular distributions are independent of the anomalous $tbW$ couplings, a
result proven earlier under less general circumstances \cite{Grzadkowski:1999iq}.
Thus, one can say that the charged-lepton angular distributions are clean and uncontaminated
probes of top polarization and thus of any new physics responsible for top production. We would
make use of this property and use charged-lepton angular distributions as a probe of top
polarization. However, to the extent that the production process also involves $tbW$ couplings,
anomalous $tbW$ couplings will enter our considerations, and we will also look at possible limits
that could be placed on them.

The rest of the paper is organized as follows. In the next section we
introduce the formalism and the framework of our work. In Section
\ref{production} we discuss the process of $Wt$ mode of single top production at the
the LHC and present our results for the observables like top polarization, charged-lepton
angular distributions and the lepton azimuthal asymmetry. Section \ref{sens} deals with the
statistical sensitivity of our observables to the anomalous couplings. The conclusions are 
given in Section \ref{conclusions}. The Appendix lists the production spin density matrix
elements at the parton level including the contributions of anomalous top couplings.

\section{Framework and analytical results}
The main aim of this work is to study the effect of anomalous top-gluon couplings on the 
top quark polarization and other angular observables in $Wt$ production at the LHC. 
For the calculation of final charged-lepton distributions, we use spin-density matrix
formalism. 
We use the narrow-width approximation (NWA) 
\beq\label{nwa}
\left|\frac{1}{p^2-m^2+im\Gamma}\right|^2\sim\frac{\pi}{m\Gamma}\delta(p^2-m^2)
\eeq
to factor the matrix amplitude squared into production and
decay parts as
\begin{eqnarray}
\overline{|{\cal M}|^2} = \frac{\pi \delta(p_t^2-m_t^2)}{\Gamma_t m_t}
\sum_{\lambda,\lambda'} \rho(\lambda,\lambda')\Gamma(\lambda,\lambda'),
\end{eqnarray}
where $\rho(\lambda,\lambda')$ and $\Gamma(\lambda,\lambda')$ are the $2 \times 2$ top
production and decay spin density matrices and $\lambda,\lambda' =\pm 1$ denotes the sign of
the top helicity. 

We obtain analytical expressions for the spin density matrix for $Wt$ production
including anomalous couplings. Use is made of the analytic manipulation program FORM 
\cite{Vermaseren:2000nd}. We find that at linear order, $\f2r$, $\rr$ and $\ri$  
give significant contributions to the production density matrix, whereas contributions from
all other couplings are proportional to the mass of $b$ quark (which we neglect consistently)
and hence vanish in the limit of zero bottom mass. To second order in anomalous couplings,
other anomalous couplings do contribute, but we focus on $\f2r$, $\rr$ and $\ri$, 
since their contributions, arising at linear order, are dominant. Expressions for the spin
density matrix elements $\rho(\pm,\pm)$ and $\rho(\pm,\mp)$ , where $\pm$ are the signs of the
top-quark helicity, for $Wt$ production including the contributions of $\rr$ and $\ri$ are given
in the Appendix. The expressions for production and top-decay spin-density matrices, for anomalous
$Wtb$ couplings, have been evaluated in Ref. \cite{Rindani:2011pk}. 

Using the NWA and spin-density matrix formalism for top production and its decay, we write the
partial cross section in the parton cm of frame as 
\bea
d\sigma&=&\frac{1}{32(2\pi)^4 \Gamma_t m_t} 
\int \left[ \sum_{\lambda,\lambda'}
\frac{d\sigma(\lambda,\lambda')}{d\cos\theta_t} \
\left(\frac{\langle\Gamma(\lambda,\lambda')\rangle}{p_t \cdot p_\ell}\right) \right] \ 
\nonumber \\
&\times& 
d\cos\theta_t \ d\cos\theta_\ell \ d\phi_\ell\ E_\ell dE_\ell \ dp_W^2,
\label{dsigell}
\eea
where the $b$-quark energy integral is replaced by an integral over the invariant mass 
$p_W^2$ of the $W$ boson, its polar-angle integral is carried out using the Dirac delta
function of Eq. (\ref{nwa}), and the average over its azimuthal angle is denoted by the angular
brackets. Integrating over the lepton energy, with limits given by 
$m_W^2<2(p_t\cdot p_\ell) < m_t^2$, the analytical expression for the differential cross section
in the parton cm frame is given as

\bea\label{angdist}
\frac{d\sigma}{d\cos\theta_t \ d\cos\theta_\ell \ d\phi_\ell} &=&  \frac{1}{32 \ \Gamma_t m_t} \ \frac{1}
{(2\pi)^4}\int \left[ \sum_{\lambda,\lambda'} \frac{d\sigma^{\lambda\lambda'}}{d \cos\theta_t} 
 g^4 \mathcal{A}^{\lambda\lambda'} \right]
|\Delta(p_W^2)|^2 dp_W^2,
\eea
where 
\bea\label{angmat1}
 \mathcal A^{\pm\pm}&=&\frac{m_t^6}{24(1-\beta_t
\cos\theta_{t\ell})^3 E_t^2}
\Big[(1-r^2)^2  (1\pm\cos\theta_{t\ell})(1\mp\beta_t)(1+2r^2)\Big],\\
\label{angmat2}
\mathcal A^{\pm\mp}&=&\frac{m_t^7}{24(1-\beta_t
\cos\theta_{t\ell})^3 E_t^3}\sin\theta_{t\ell} e^{\pm i\phi_\ell}
\Big[(1-r^2)^2(1+2r^2)\Big].
 \eea
Here $r=m_W/m_t$ and $\cos\theta_{t\ell}$ is the angle between the top
quark and the charged lepton in top decay in the parton cm frame, 
given by
\begin{equation}
 \cos \theta_{t \ell}=\cos \theta_t \cos \theta_{\ell}+\sin \theta_t \sin \theta_{\ell} \cos \phi_{\ell},
\label{costhetatl}
\end{equation}
where $\theta_\ell$ and $\phi_\ell$ are the lepton polar and azimuthal angles.

\section{Single-top production in association with a $W$ boson}\label{production}

\begin{figure}[h!]
\begin{center}
 \includegraphics[scale=0.35]{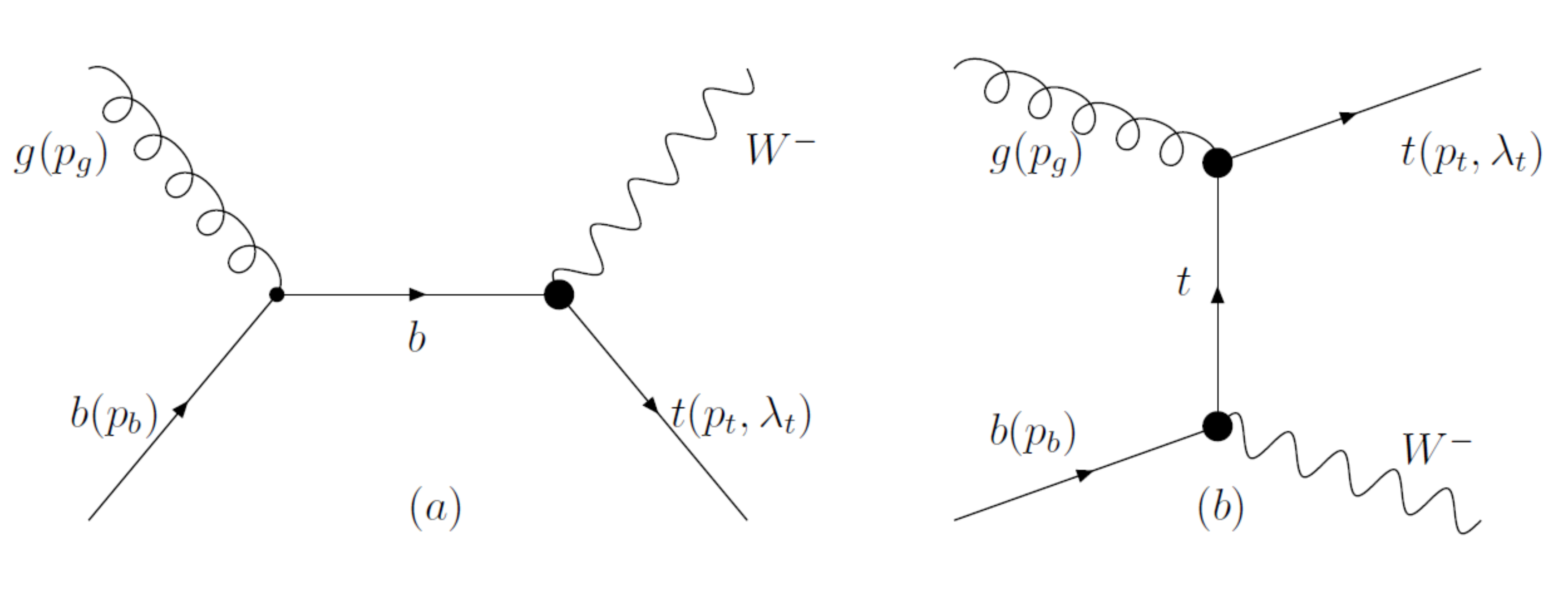} 
\end{center}
\caption{Feynman diagrams for the $Wt$ production process. 
The blobs here denote the anomalous $tbW$ and $ttg$ couplings.} 
\label{feyngraph1}
\end{figure}

The theoretical mechanism for the $tW^-$ mode of single-top production has been studied in detail in Refs. 
\cite{Tait:1999cf,White:2009yt,Frixione:2008yi}.
At the parton level, the $tW^-$ production proceeds through a gluon and a bottom quark each
coming from a proton and gets contributions from two diagrams. Feynman diagrams  for the
process $g(p_g)b(p_b)\rightarrow t(p_t,\lambda_t)W^-$, where $\lambda_t=\pm1$ represents the top
helicity, are shown in Fig. \ref{feyngraph1}. The blobs denote effective $tbW$ and $ttg$
vertices, including anomalous couplings, in the production process. 

\begin{figure}[h!]
\includegraphics[scale=0.7]{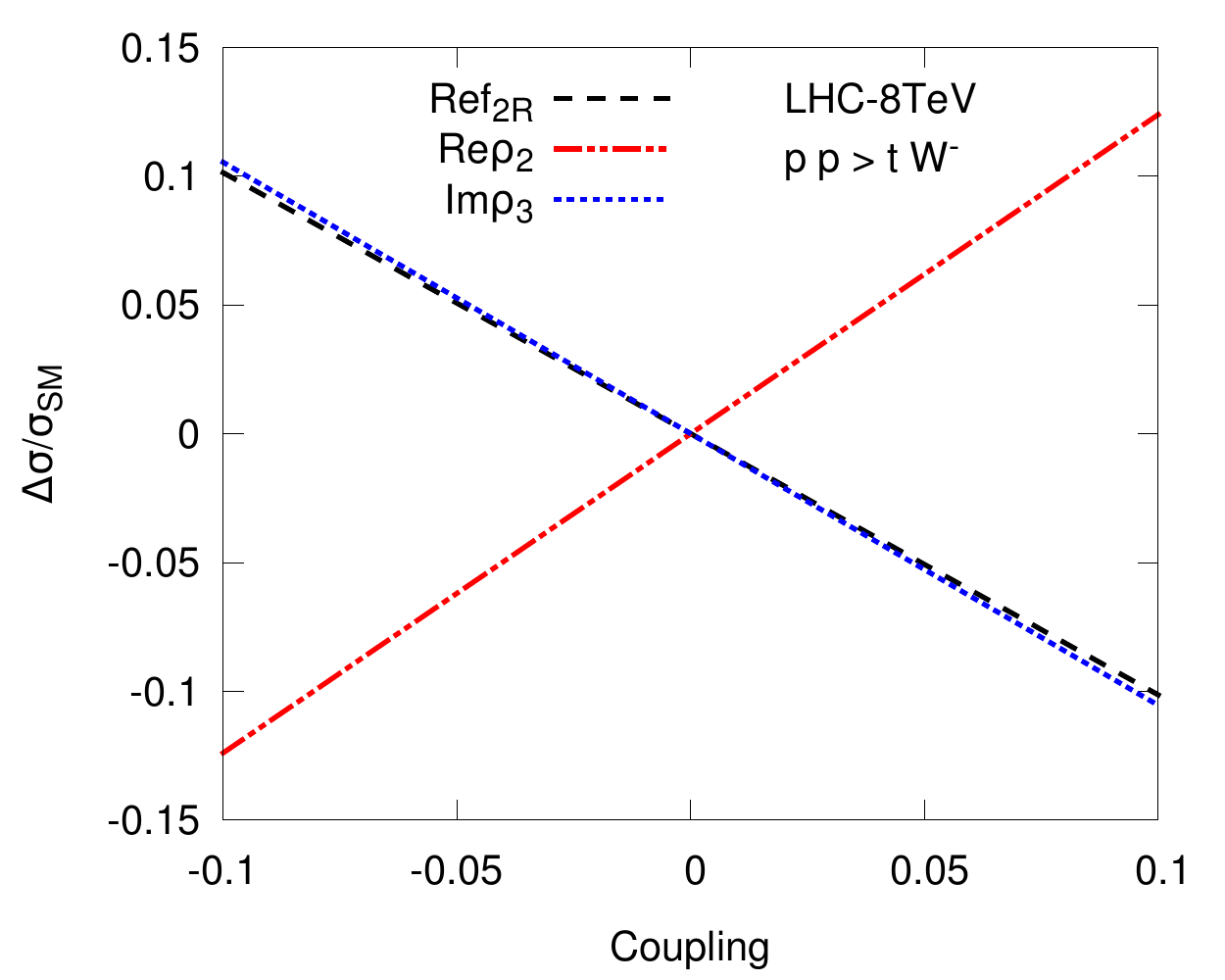} 
\includegraphics[scale=0.7]{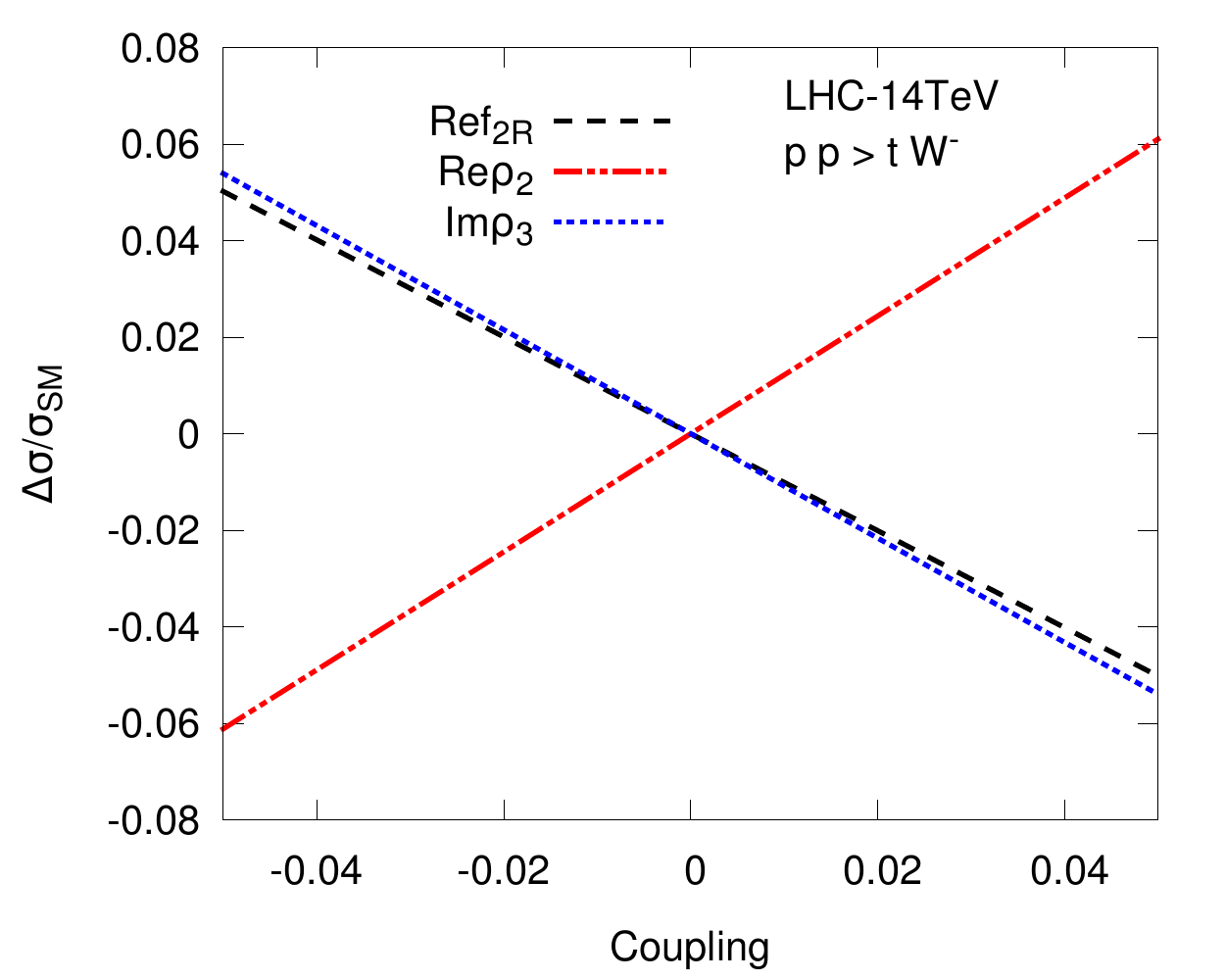} 
\caption{ The cross section for $tW^-$ production at the LHC8 (left) and LHC14 (right), 
as a function of the anomalous $tbW$ and $ttg$ couplings with linear approximation.} 
\label{cs-lin}
\end{figure}

For numerical calculations, we use the leading-order parton distribution function (PDF) sets of
CTEQ6L \cite{cteq6}, with a factorization scale of $m_t=173.2$ GeV. We also evaluate the strong
coupling at the same scale, $\alpha_s(m_t)=0.1085$. We make use of the following values of other
parameters: $M_W=80.403$ GeV, the electromagnetic coupling $\alpha_{em}(m_Z)=1/128$ and 
$\sin^2\theta_W=0.23$. We set $\fla = 1$, $\rho_1=1$ and $V_{tb}=1$ in our calculations. 
We take only one coupling to be non-zero at a time in the analysis, except in Sec.\ref{sens}.

\begin{figure}[h!]
\includegraphics[scale=0.7]{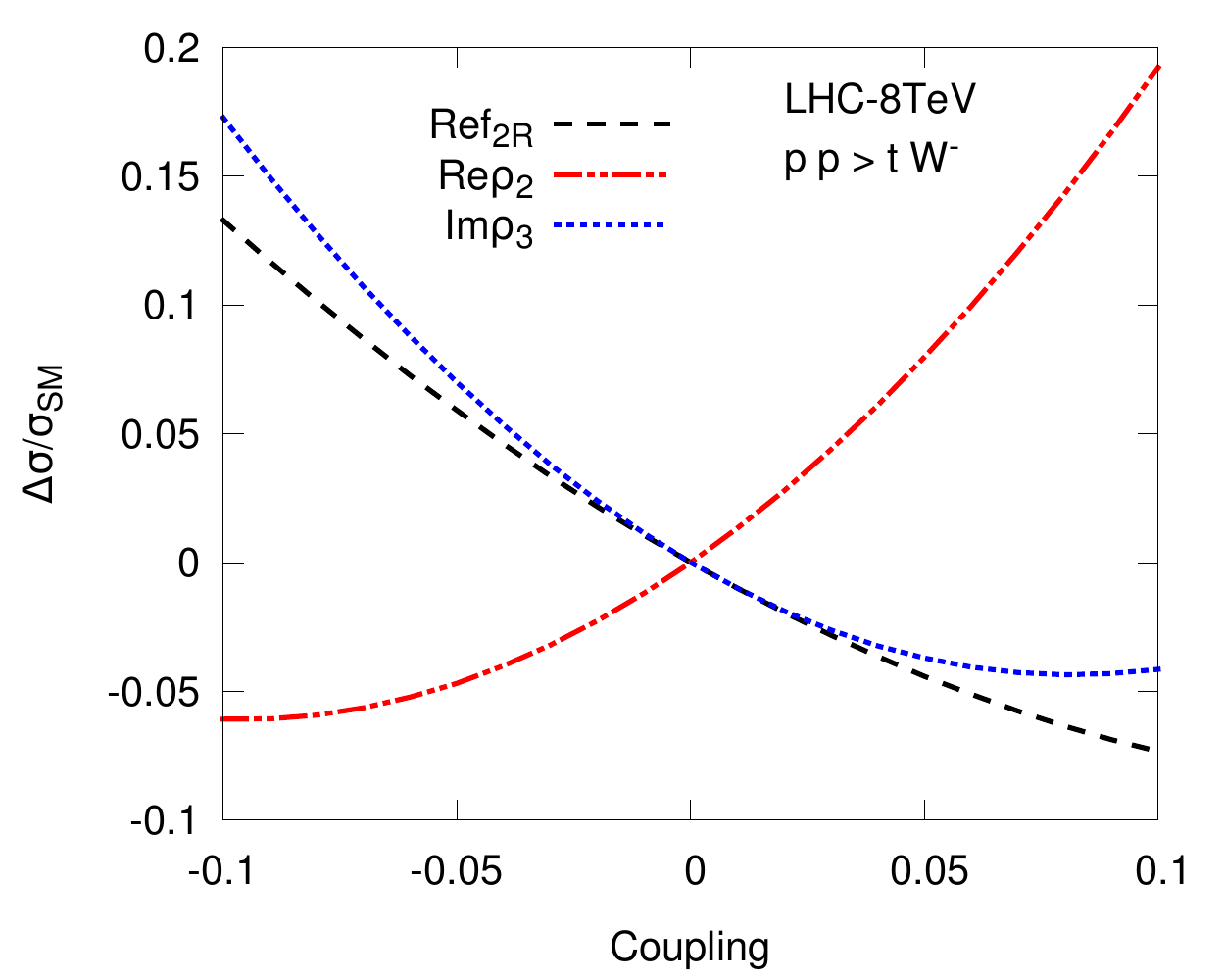} 
\includegraphics[scale=0.7]{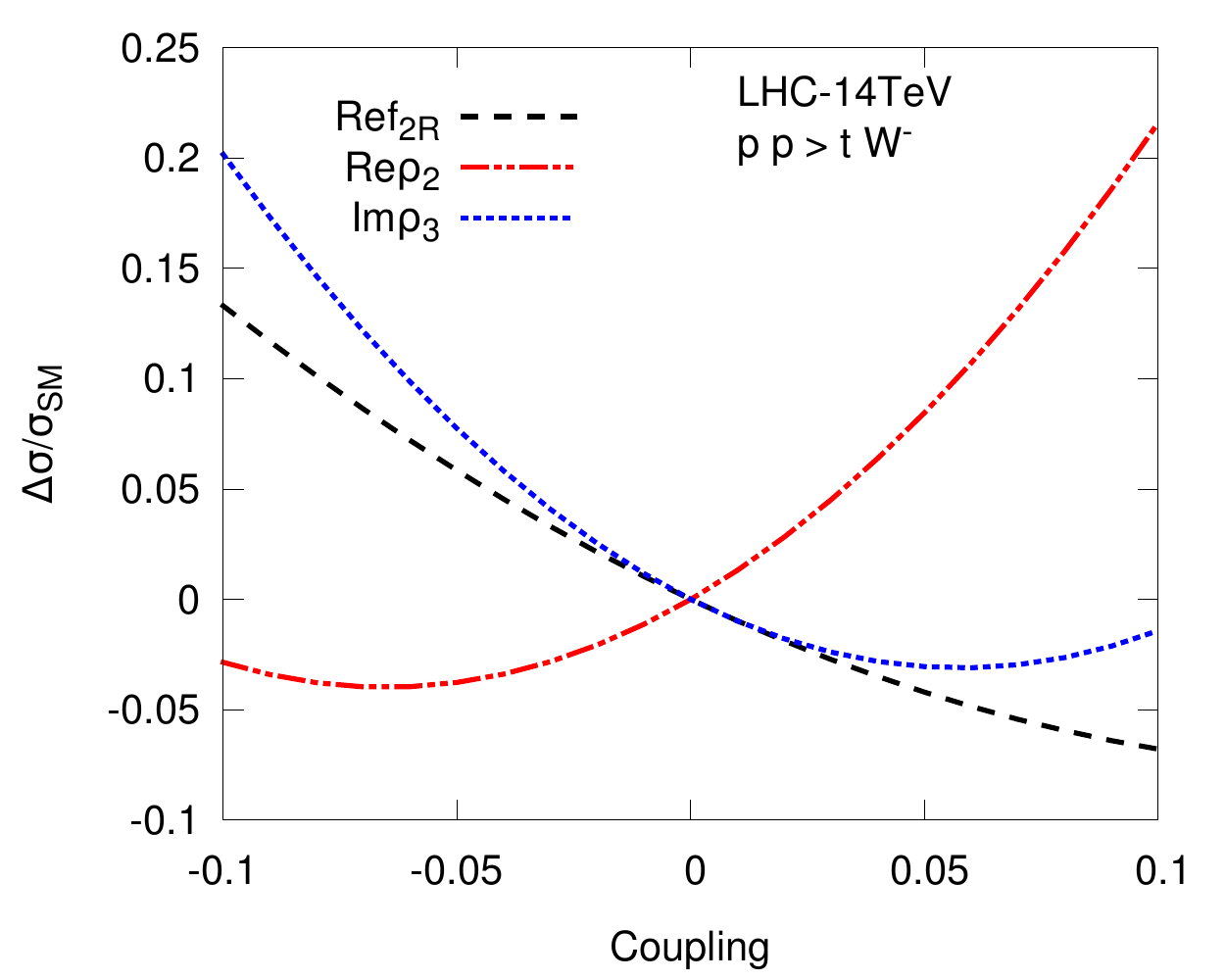} 
\caption{ The cross section for $tW^-$ production at the LHC8 (left) and LHC14 (right), as a 
function of the anomalous $tbW$ and $ttg$ couplings including their contributions at all orders.} 
\label{cs-full}
\end{figure}

After integrating the density matrix given in the Appendix over the phase space, the diagonal
elements of this integrated density matrix, which we denote by $\sigma(+,+)$ and  $\sigma(-,-)$,
are respectively the cross sections for the production of positive and negative helicity tops and
$\sigma_{\rm{tot}}=\sigma(+,+)+\sigma(-,-)$ is the total cross section. 
We include the contributions of the anomalous couplings to the cross section at linear order, as 
well as without that approximation. Since the cross section may receive large radiative 
corrections at the LHC, we focus on using observables like asymmetries which are ratios 
of some partial cross sections and are expected to be insensitive to such corrections. 

In Fig. \ref{cs-lin}, we show the relative change in cross section, $\Delta\sigma/\sigma_{SM}$ as
a function of various anomalous $tbW$ and $ttg$ couplings when their contributions are taken upto
linear order for LHC8 (left) and LHC14 (right), while in Fig. \ref{cs-full} corresponding plots
are shown for the full contributions of the anomalous couplings for LHC8 (left) and LHC14(right).
From Fig. \ref{cs-full}, one can infer that the cross section is very
sensitive to negative values of $\f2r$ and $\ri$ and for positive values of $\rr$. The linear
approximation is seen to be good for values of anomalous couplings $\f2r$ and $\rr$ ranging from
$-0.05$ to 0.05 while for $\ri$ it is valid only in the range [-0.03,0.03]. 

Recently, the ATLAS and CMS collaborations reported a combination of cross section measurements
for $tW$ single top production at $\sqrt{s}=8$ TeV with integrated luminosities of 20.3 fb$^{-1}$ 
and 12.2 fb$^{-1}$, respectively, to be $25.0\pm 4.7$ pb \cite{combined}, in agreement with the SM
prediction. Using the result of ref. [34] which provides a combined value based
upon the cross section measurements from ATLAS and CMS with experimental error, we determine that the fractional 
change in cross section, at the level of 1$\sigma$, constrains the allowed range for $\rr$ at [-0.3,~+0.1] 
and that for $\ri$ at [-0.1,~ +0.3]. The limits can be improved
at 14 TeV only with much larger integrated luminosity. Ref. \cite{Ayazi:2013cba} estimates
that cross section measurement can give constraints on $\rr$ and $\ri$ of the order of $\pm 0.01$
and $\pm 0.02$ at the LHC14, assuming 5\%  uncertainty in the measurement. The SM cross section, as also the 
cross section including anomalous couplings used for the determination of these constraints, are from our computation at the leading order. 
However, since we consider a fractional change in the cross section to derive our limits, 
it would be expected to be stable to higher-order corrections and also other uncertainties 
like PDF and scale uncertainties.

\subsection{Top angular distribution}

The angular distribution of the top quark would be modified by anomalous couplings. Since the top
quark is produced in a $2\to 2$ process, its azimuthal distribution is flat. We can study its
polar distribution with the polar angle defined with respect to either of the beam directions as
the $z$ axis. We find that the polar distribution is sensitive to anomalous $tbW$ couplings.

\begin{figure}[h!]
\includegraphics[scale=0.7]{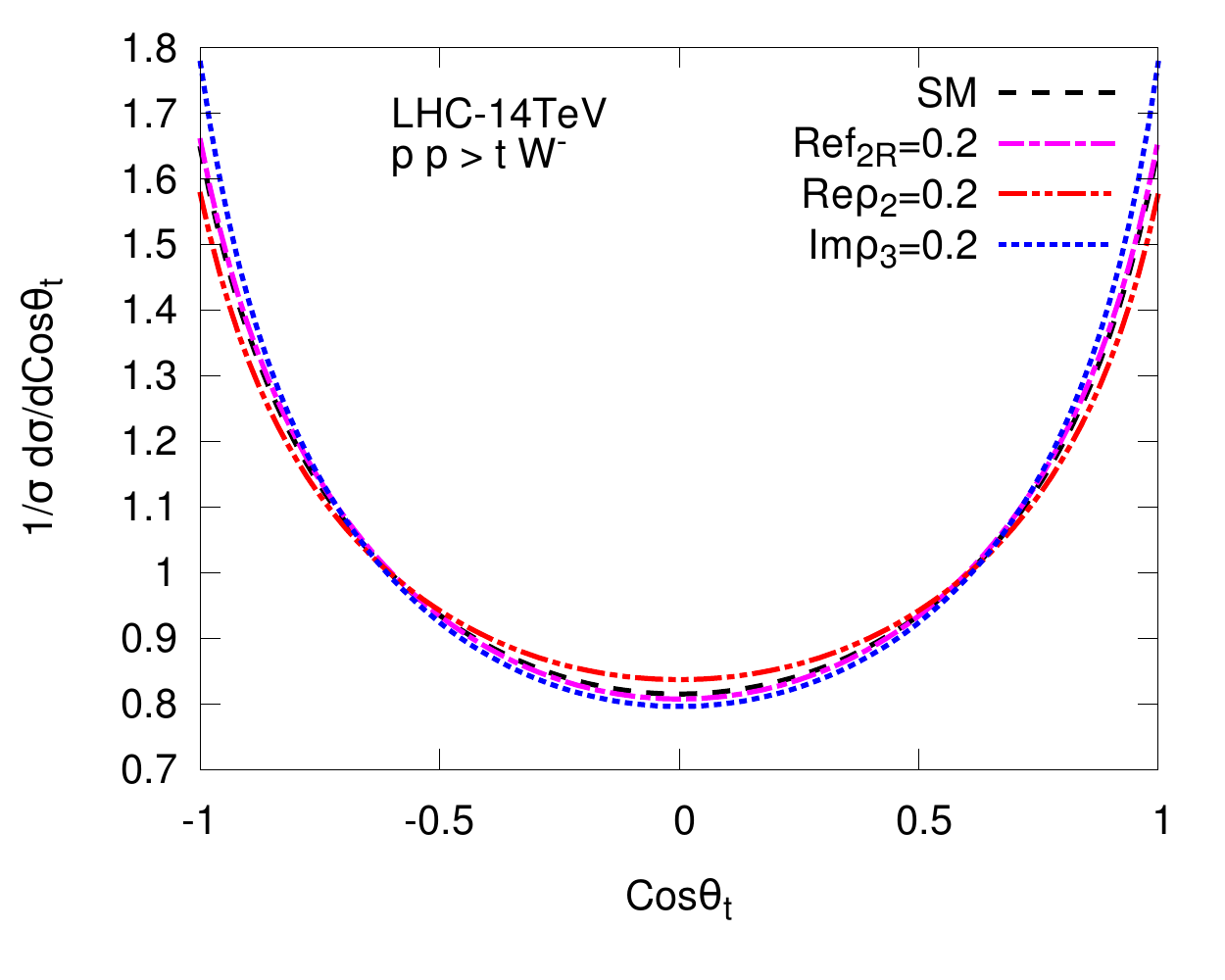} 
\includegraphics[scale=0.7]{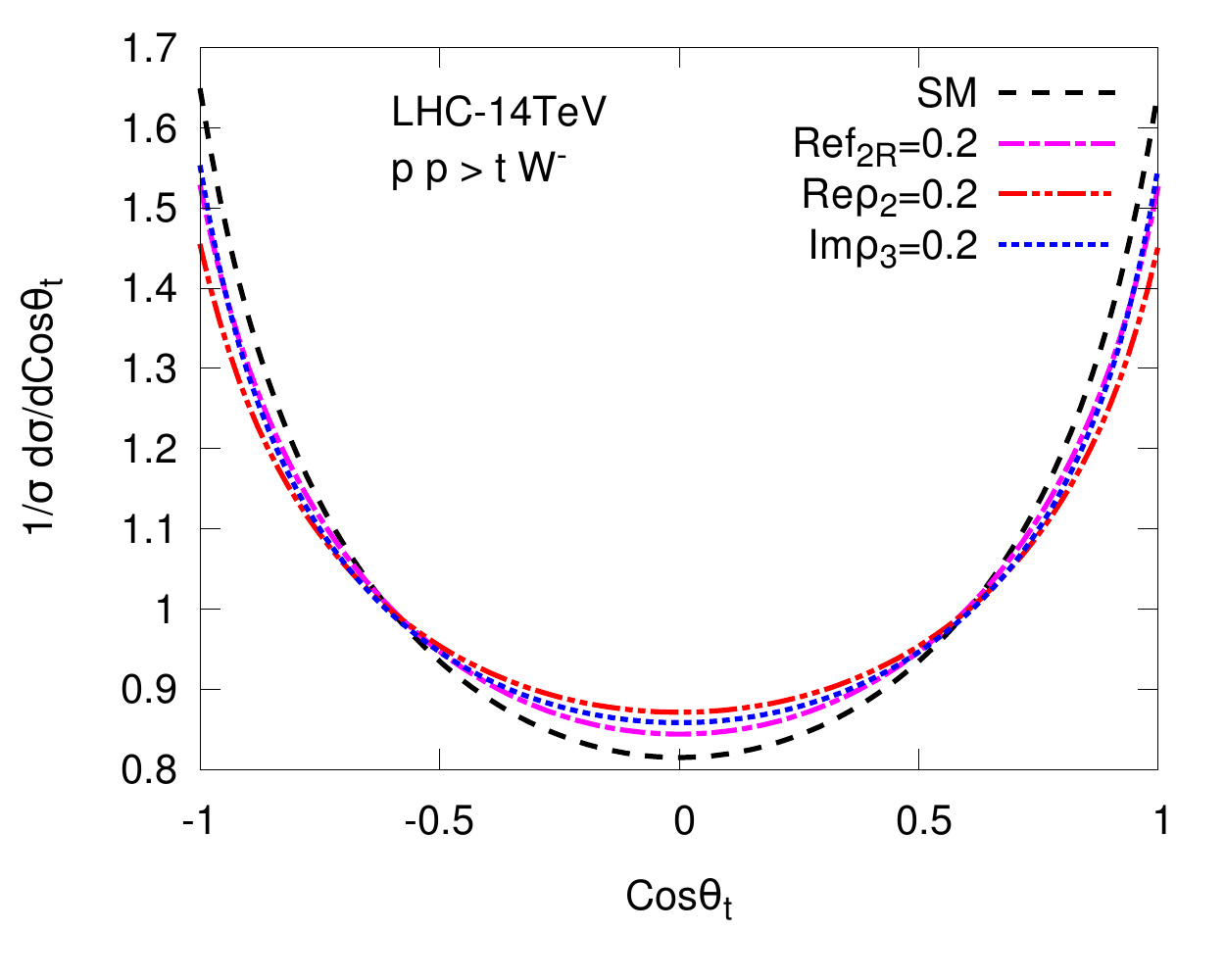} 
\caption{ The top polar angular distributions for $tW^-$ production at the LHC14 
for different anomalous $tbW$ and $ttg$ couplings with linear approximation (left) and 
without that approximation (right).} 
\label{polar-dist-top}
\end{figure}

\begin{figure}[h!]
\includegraphics[scale=0.7]{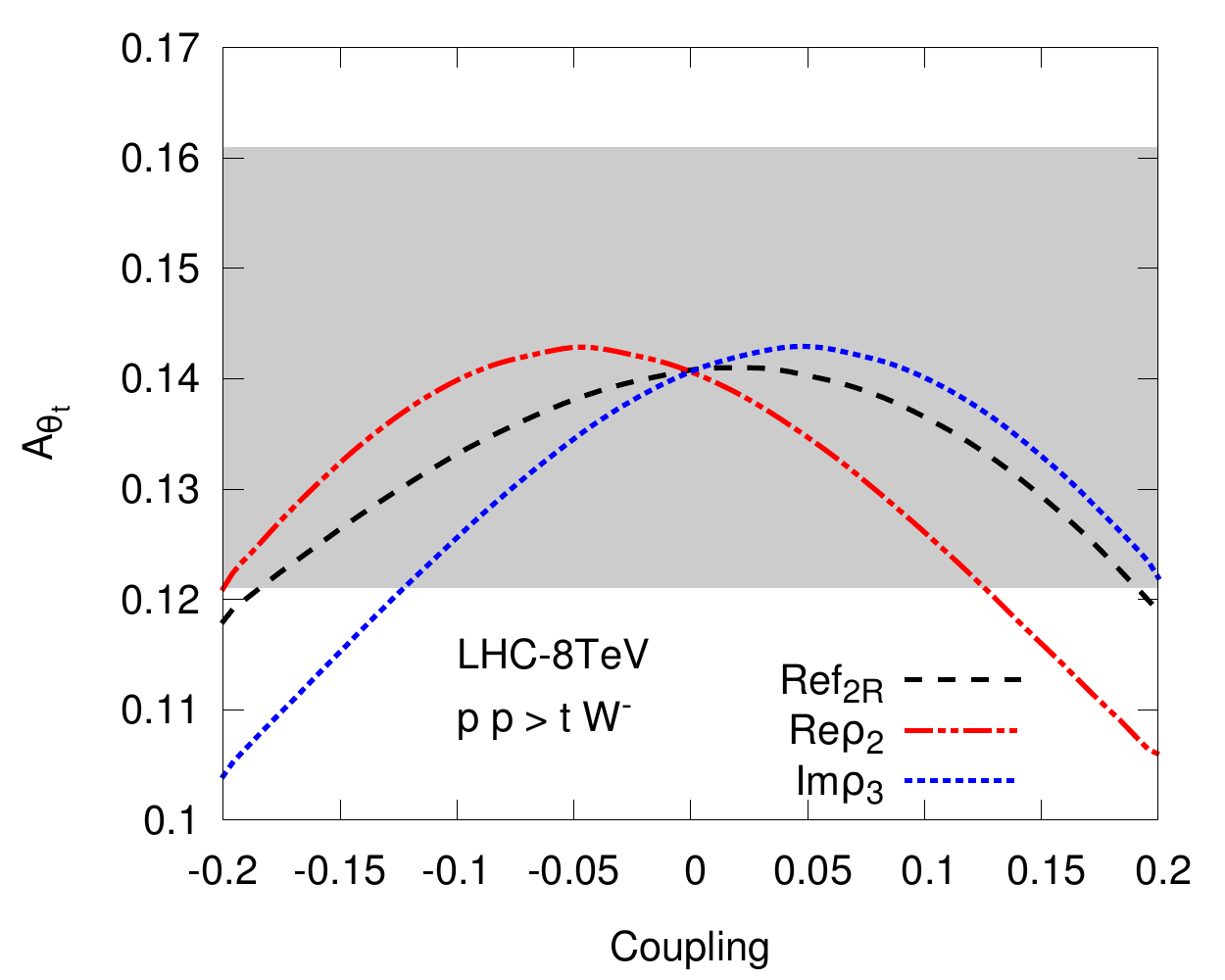} 
\includegraphics[scale=0.7]{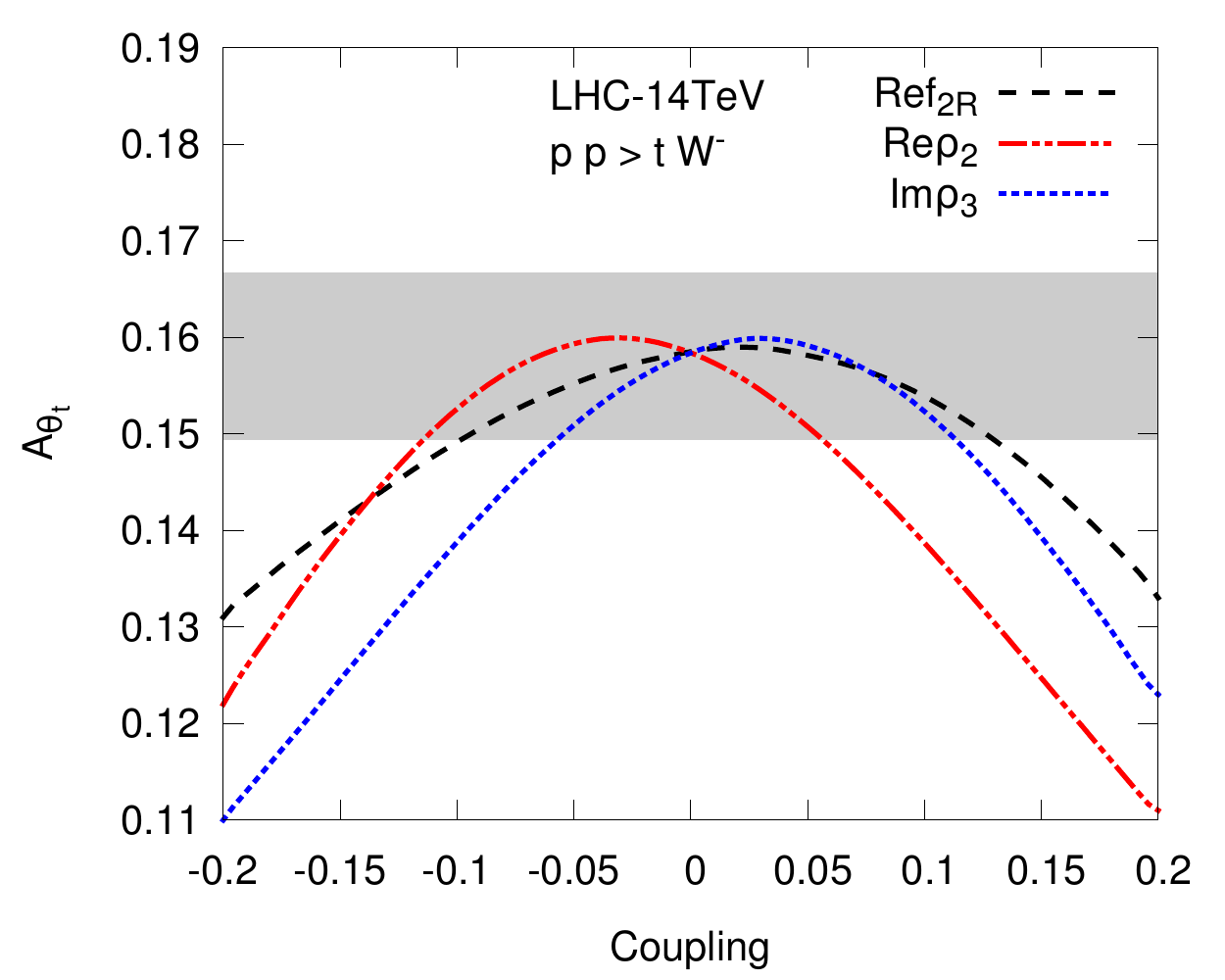} 
\caption{ The top polar asymmetries for $tW^-$ production at the LHC8 (left) and LHC14 (right), 
as a function of anomalous $tbW$ and $ttg$ couplings. The grey band corresponds to the top polar asymmetry predicted in the SM with a 1 $\sigma$ error interval. } 
\label{asym-top}
\end{figure}

The normalized polar distribution is plotted in Fig. \ref{polar-dist-top} for LHC8 and LHC14.
As can be seen from Fig. \ref{polar-dist-top}, the curves for the polar distributions for the SM
and for the anomalous couplings of magnitude $0.2$ are separated from each other. As the 
colliding beams are identical in the case for LHC, the top polar distribution has no forward
-backward asymmetry. However, we can define an asymmetry utilizing the polar distributions of 
the top quark as 
\beq
\mathcal A_{\theta}^t=\frac{\sigma(|z|>0.5)-\sigma(|z|<0.5)}{\sigma(|z|>0.5)+\sigma(|z|<0.5)}
\eeq
where $z$ is $\cos\theta_t$. We plot this asymmetry as a function of anomalous $tbW$ and $ttg$
couplings for LHC8 and LHC14 in Fig. \ref{asym-top}. 

The asymmetry $\mathcal A_{\theta}^t$ requires accurate determination of the top direction in
the lab frame and a quantitative estimate of its sensitivity to anomalous couplings needs details
of the efficiency of reconstruction of the direction. 
We do not study this asymmetry any further, but proceed to a discussion of top polarization. 

\subsection{Top polarization}

\begin{equation}
P_t=\frac{\sigma(+,+)-\sigma(-,-)}{\sigma(+,+)+\sigma(-,-)}.
\label{eta3def}
\end{equation}
\begin{figure}[h!]
\includegraphics[scale=0.7]{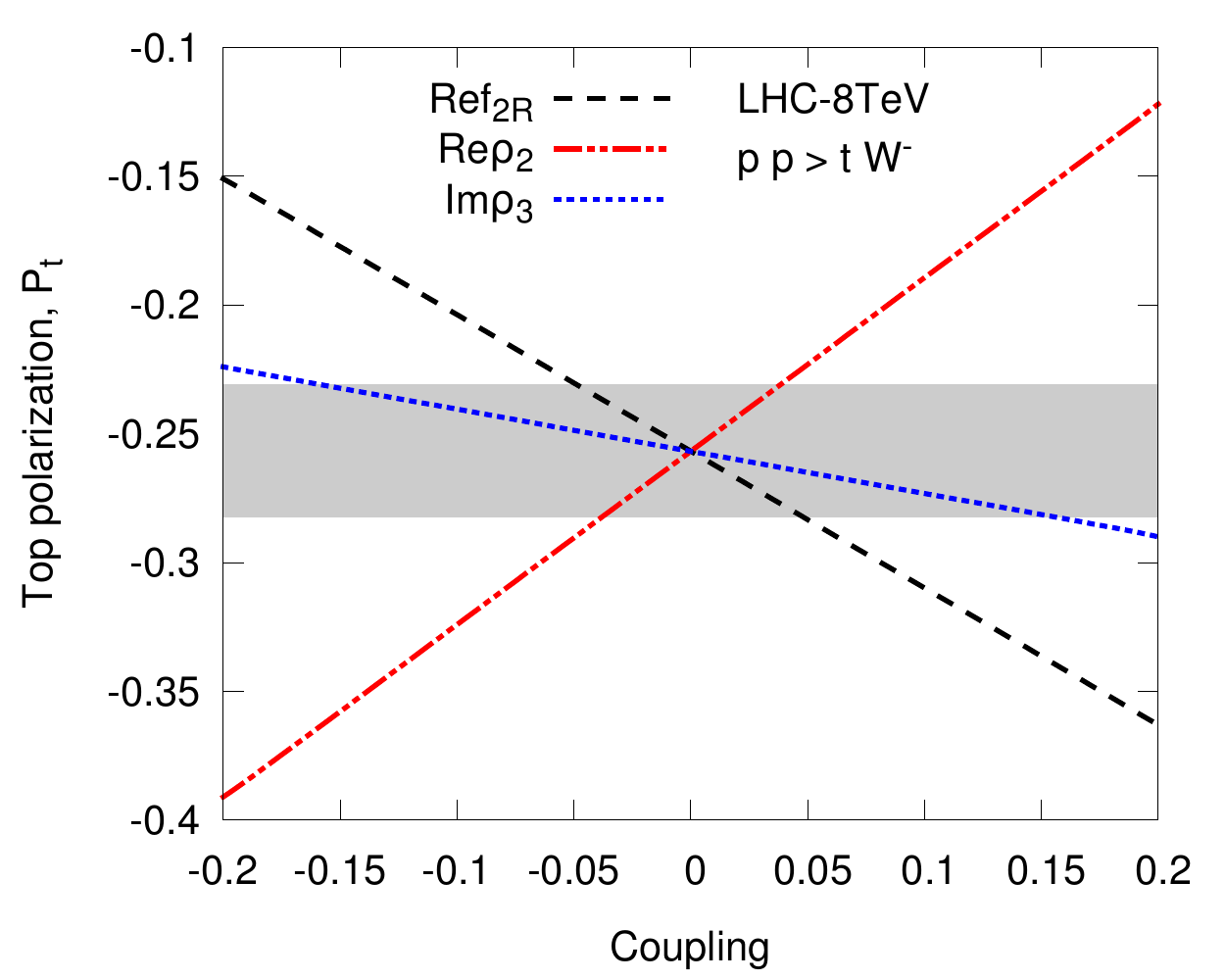} 
\includegraphics[scale=0.7]{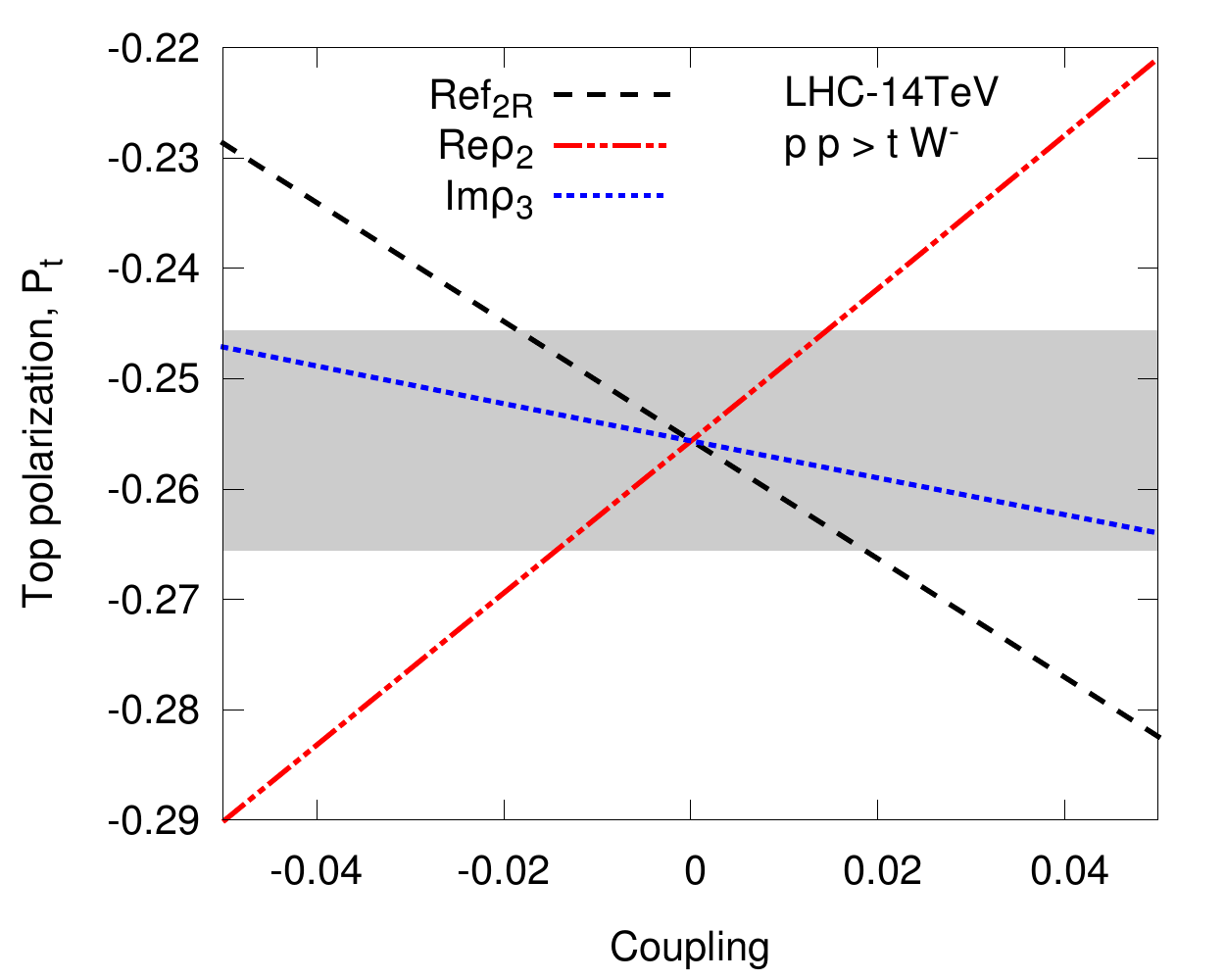} 
\caption{ The top polarization for $tW^-$ production at the LHC8 (left) and LHC14
(right) as a function of the anomalous $tbW$ and $ttg$ couplings. The grey band corresponds to the
top polarization predicted in the SM with a 1 $\sigma$ error interval. } 
\label{polcoup-lin}
\end{figure}

\begin{figure}[h!]
\includegraphics[scale=0.7]{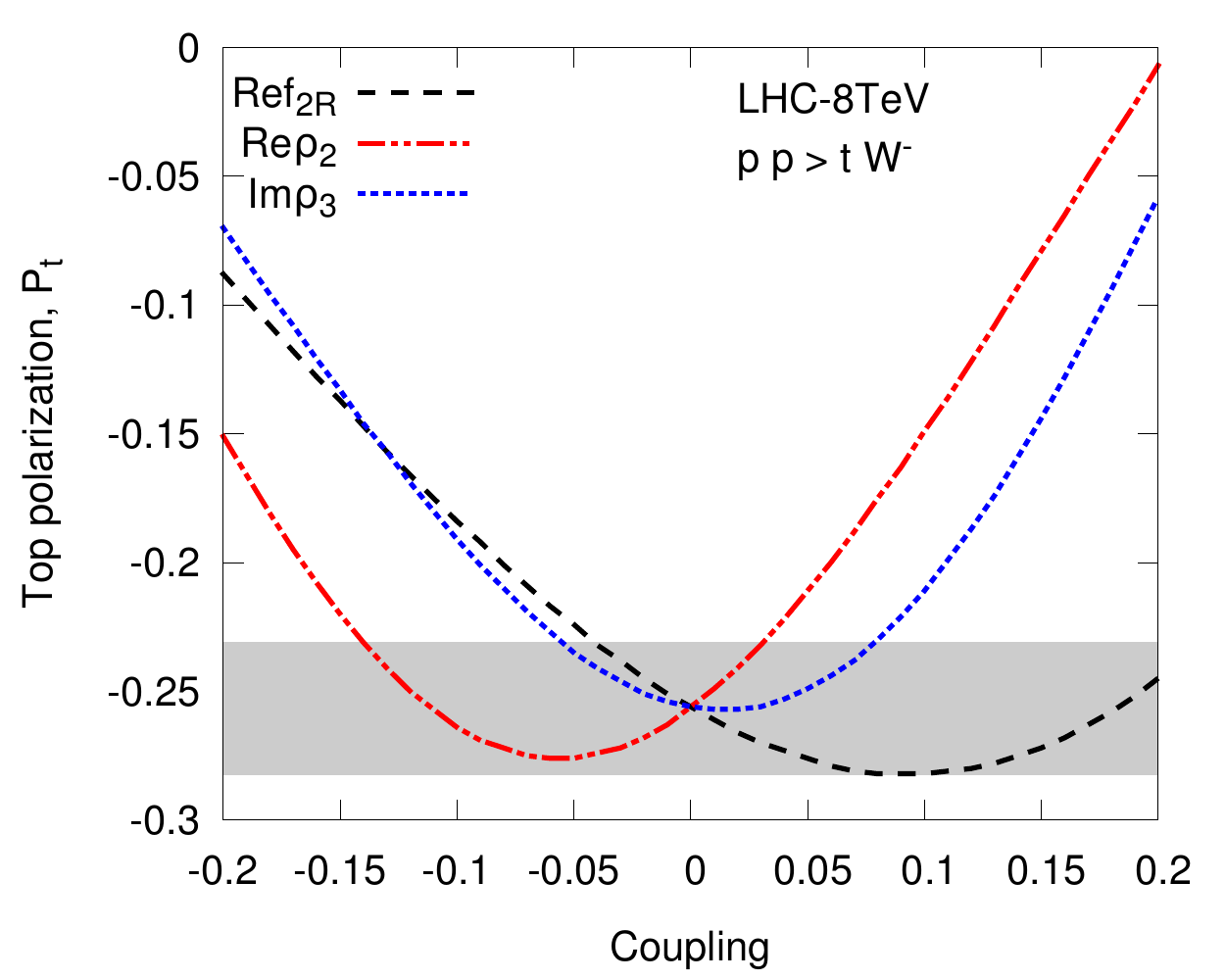} 
\includegraphics[scale=0.7]{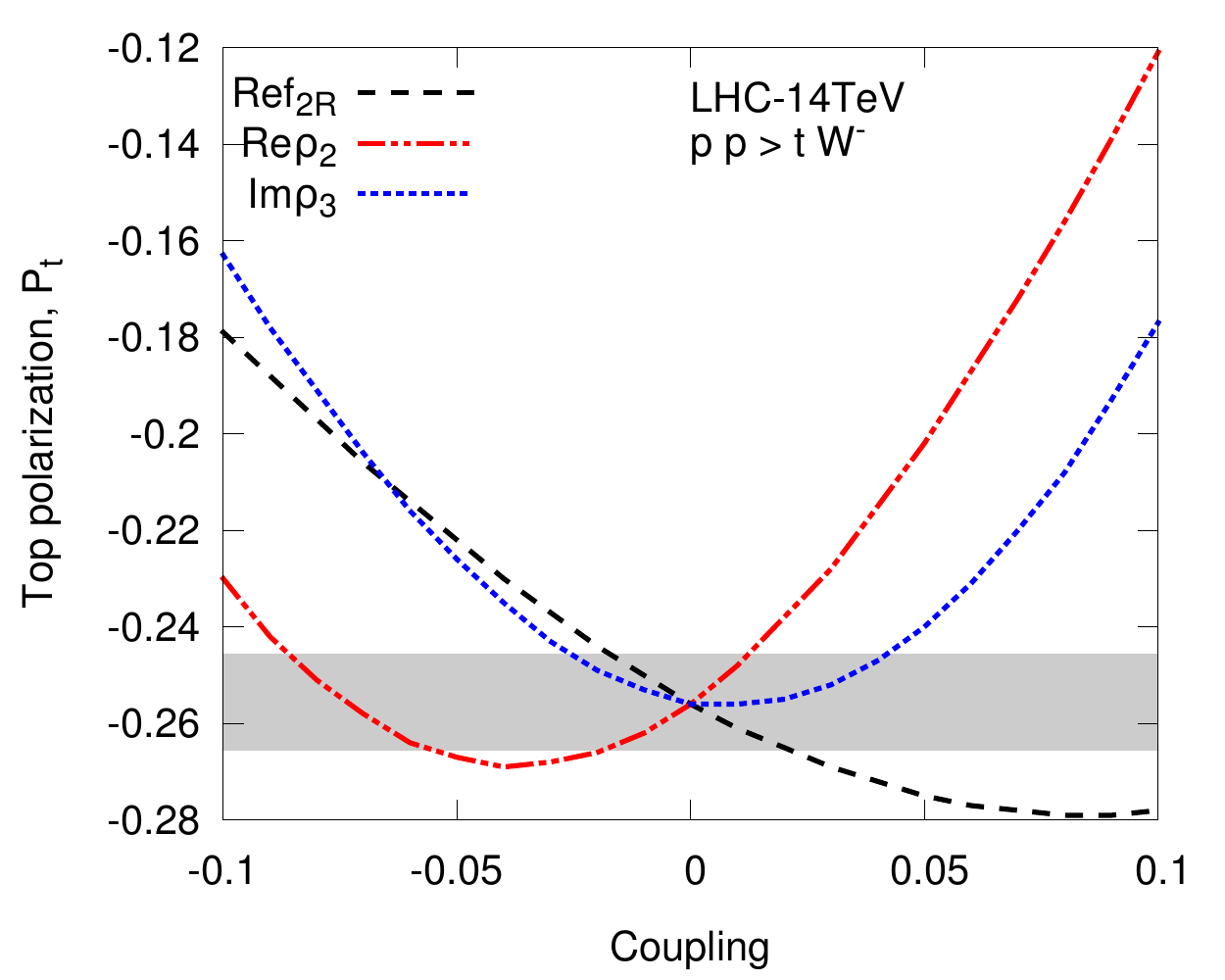} 
\caption{ The top polarization for $tW^-$ production at the LHC8 (left) and LHC14 (right) as a function of anomalous $tbW$ and $ttg$ 
couplings. The grey band corresponds to the top polarization predicted in the SM with a 1 $\sigma$ error interval. } 
\label{polcoup-full}
\end{figure}

\noindent This polarization is shown in Fig. \ref{polcoup-lin} as a function of the anomalous
couplings in the linear approximation for LHC8 and LHC14. In Fig. \ref{polcoup-full} we show the
top polarization when full contributions of the couplings have been included. As compared to the
SM value of $-0.26$ for $\sqrt{s}=14$ TeV, the degree of longitudinal top polarization varies
from $-0.18$ to $-0.28$ for $\f2r$ and $-0.27$ to $-0.12$ for $\rr$ varied over the range $-0.1$
to $+0.1$, while it varies from $-0.16$ to $-0.26$ for the same range of $\ri$ and is almost
symmetric about $\ri=0$, which signifies a very small contribution at linear order for $\ri$.

We notice that, just as for the total cross section, $P_t$ is more sensitive to negative values of
$\f2r$ while for $\rr$ it is more sensitive to positive values of the coupling. Thus $P_t$ will be
a very good probe of $\f2r$, $\rr$ and $\ri$ if it can be measured at the LHC. However, the standard measurement of 
$P_t$ requires reconstruction of the top rest frame, which is a difficult
task, and would entail a reduction in efficiency. We will therefore investigate lab frame decay distributions for the measurement of the anomalous couplings. 

All the quantities considered so far, viz., the total cross section,and the top polarization, 
can only be measured using information from the decay of the top. Both the polar distribution and
the top polarization would play a role in determining the distributions of the decay products. 
Our main aim is to devise observables in single-top production which can be measured in the lab
frame and give a good estimate of top polarization and hence probe anomalous $ttg$ couplings.
We proceed to construct such observables from the kinematic variables of the charged lepton  
produced in the decay of the top.

\subsection{Angular distributions of the charged lepton}
Top polarization can be determined through the angular distribution of its decay products. In the SM, the dominant decay mode is $t\to b W^+$, 
with a branching ratio (BR) of 0.998, with the $W^+$ subsequently decaying to $\ell^+ \nu_\ell$ (semileptonic decay, BR 1/9 for each lepton) or 
$u \bar{d}$, $c\bar{s}$ (hadronic decay, BR 2/3). The angular distribution of a decay product $f$ for a top-quark ensemble has the form 
 \begin{equation}
 \frac{1}{\Gamma_f}\frac{\textrm{d}\Gamma_f}{\textrm{d} \cos \theta _f}=\frac{1}{2}(1+\kappa _f P_t \cos \theta _f).
 \label{topdecaywidth}
 \end{equation}
 Here $\theta_f$ is the angle between the momentum of fermion $f$ and the top spin vector in the
 top rest frame and $P_t$ (defined in Eq. (\ref{eta3def})) 
is the degree of polarization of the top-quark ensemble. $\Gamma_f$ is the partial decay width and $\kappa_f$ is the spin analyzing power of $f$. 
Obviously, a larger $\kappa_f$ makes $f$ a more sensitive probe of the top spin. The charged lepton and the $d$ quark are the best spin analyzers 
with $\kappa_{\ell^+}=\kappa_{\bar{d}}=1$, while $\kappa_{\nu_\ell}=\kappa_{u}=-0.30$ and $\kappa_{b}=-\kappa_{W^+}=-0.39$, all $\kappa$ 
values being at tree level \cite{Bernreuther:2008ju}. Thus the $\ell^+$ or $d$ have the largest probability of being emitted in the 
direction of the top spin and the least probability in the direction opposite to the spin. Since at the LHC, the lepton energy and 
momentum can be measured with high precision, we focus on leptonic decays of the top. 

\begin{figure}[h!]
\begin{center}
\includegraphics[scale=0.65]{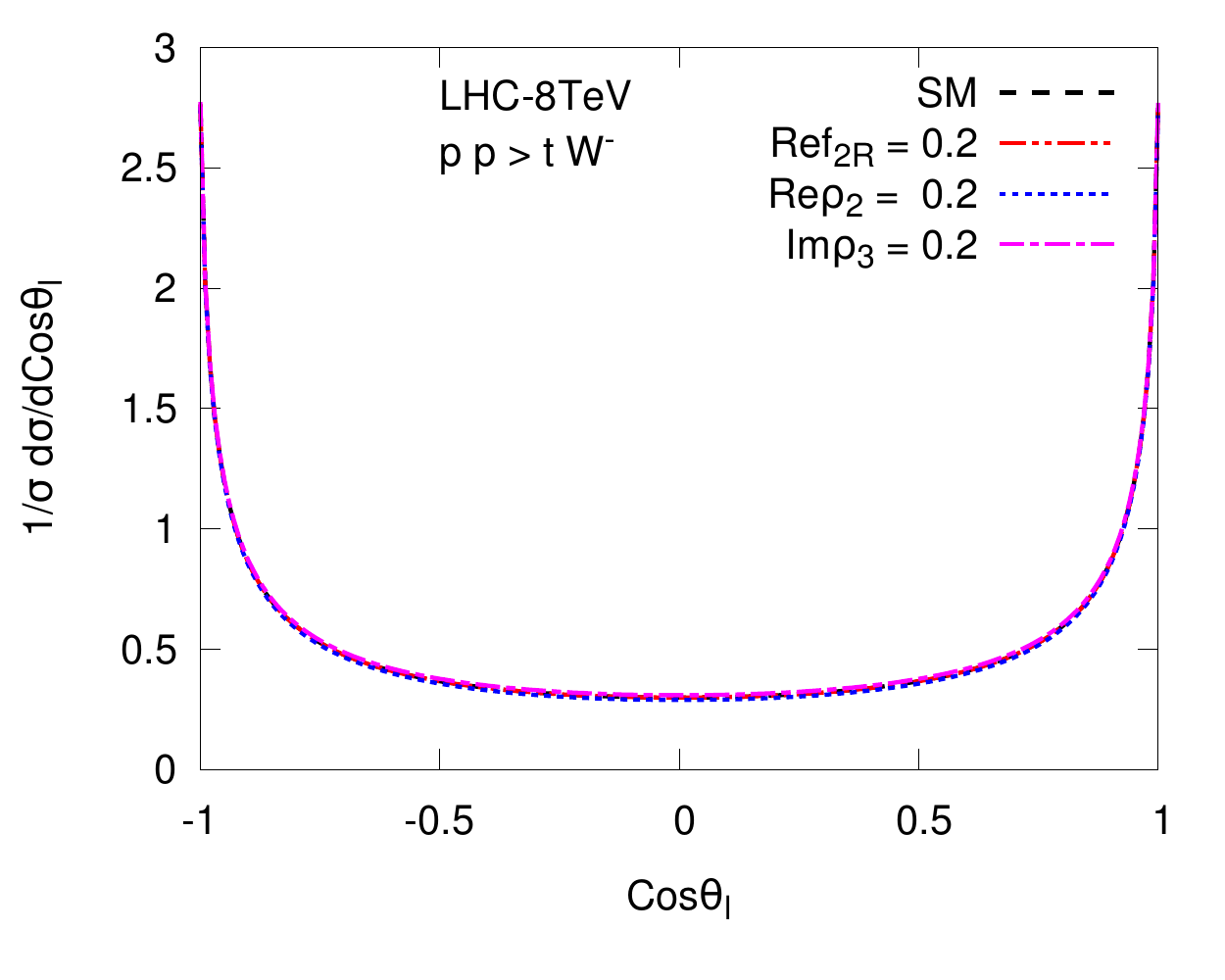}
\includegraphics[scale=0.65]{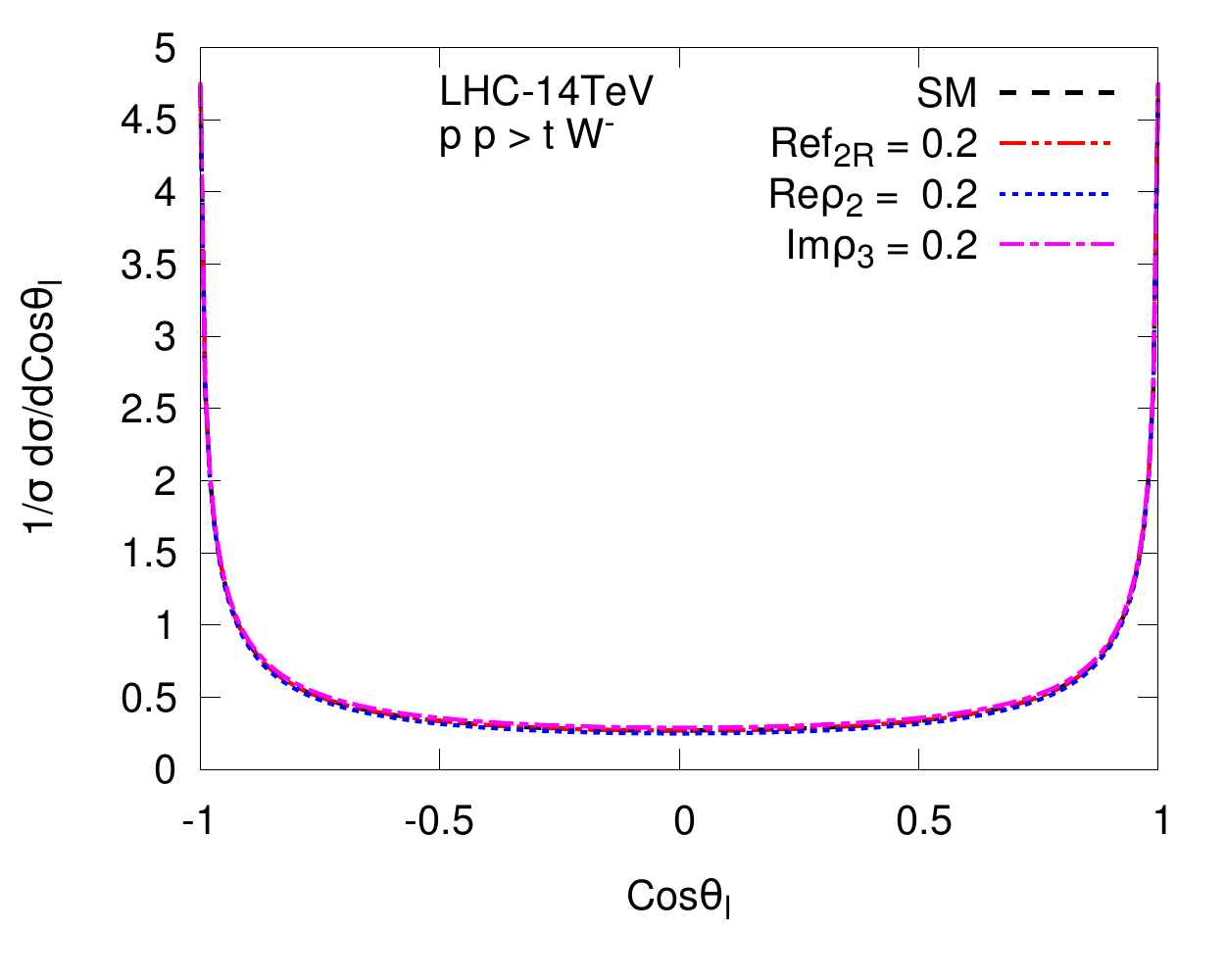}
\caption{ The normalized polar-angle distribution of the charged lepton in associated-$W$t
single-top production at the LHC8 (left) and LHC14 (right) for the SM and with anomalous 
$Wtb$ and $ttg$ couplings.} 
\label{dist-polar}
\end{center}
\end{figure}

To reconstruct the top-rest frame, one needs full 
information about the top momentum. However, because of the missing neutrino, it
is not possible to  reconstruct  completely and unambiguously 
the top longitudinal momentum. This incomplete information may lead to large
systematic errors. In this work, we focus on laboratory-frame angular distributions of the
charged lepton and thus do not require a full determination of the top momentum. In this sense, 
the observables we construct are more robust against systematic errors.
Also, as mentioned earlier and shown in Refs. \cite{Godbole:2006tq}, the charged-lepton
angular distribution in the lab frame is independent of  any new physics in top decay and  is
thus a clean and uncontaminated probe of new physics in top production.

In the lab frame, we define the lepton polar angle w.r.t. either beam direction as the $z$
axis and the azimuthal angle with respect to the top-production plane chosen, as the $x$-$z$ plane,
with the convention that the $x$ component of the top momentum is positive. At the LHC, which is a 
symmetric collider, it is not possible to define a positive sense for the $z$ axis. Hence the
lepton angular distribution is symmetric under interchange of $\theta_{\ell}$ and 
$\pi-\theta_{\ell}$ as well as of $\phi_{\ell}$ and $2\pi-\phi_{\ell}$.

We first look at the polar-angle distribution of the charged lepton and  the effect on it of 
anomalous $Wtb$ and $ttg$ couplings. As can be seen from Fig. \ref{dist-polar}, where we plot
the polar-angle distribution for LHC8 and LHC14, the normalized distributions are insensitive
to anomalous $ttg$ couplings. The sensitivity of polar-angle distributions on the anomalous
$ttg$ couplings in top-pair production have been studied in detail in Ref. 
\cite{hioki, Biswal:2012dr} for the Tevatron, LHC7 and LHC14 where they find that the polar 
distribution of the charged lepton is quite sensitive at the Tevatron to anomalous $ttg$ couplings but much less sensitive at the LHC in pair production of top quarks.

\begin{figure}[h!]
\begin{center}
\includegraphics[scale=0.65]{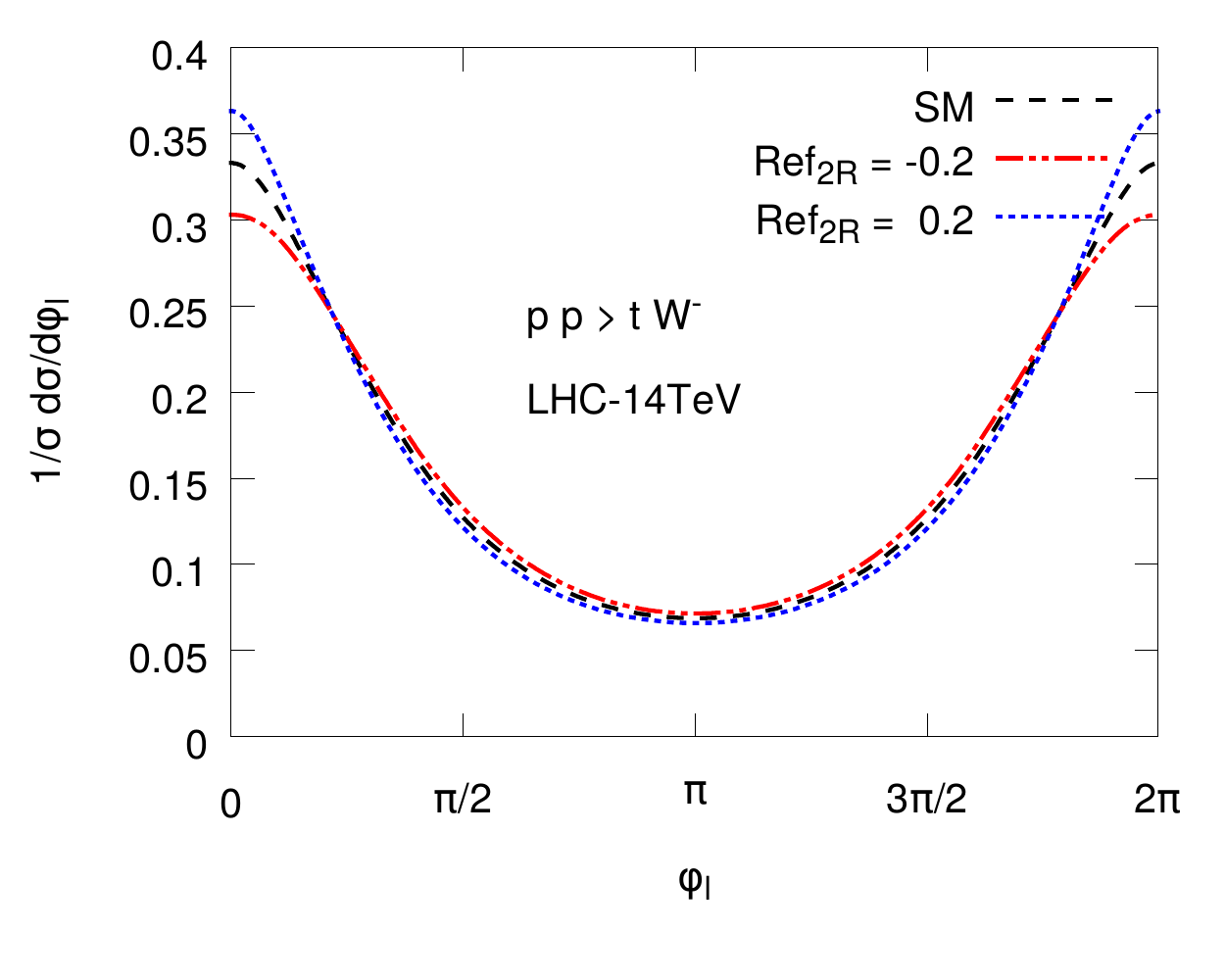}
\includegraphics[scale=0.65]{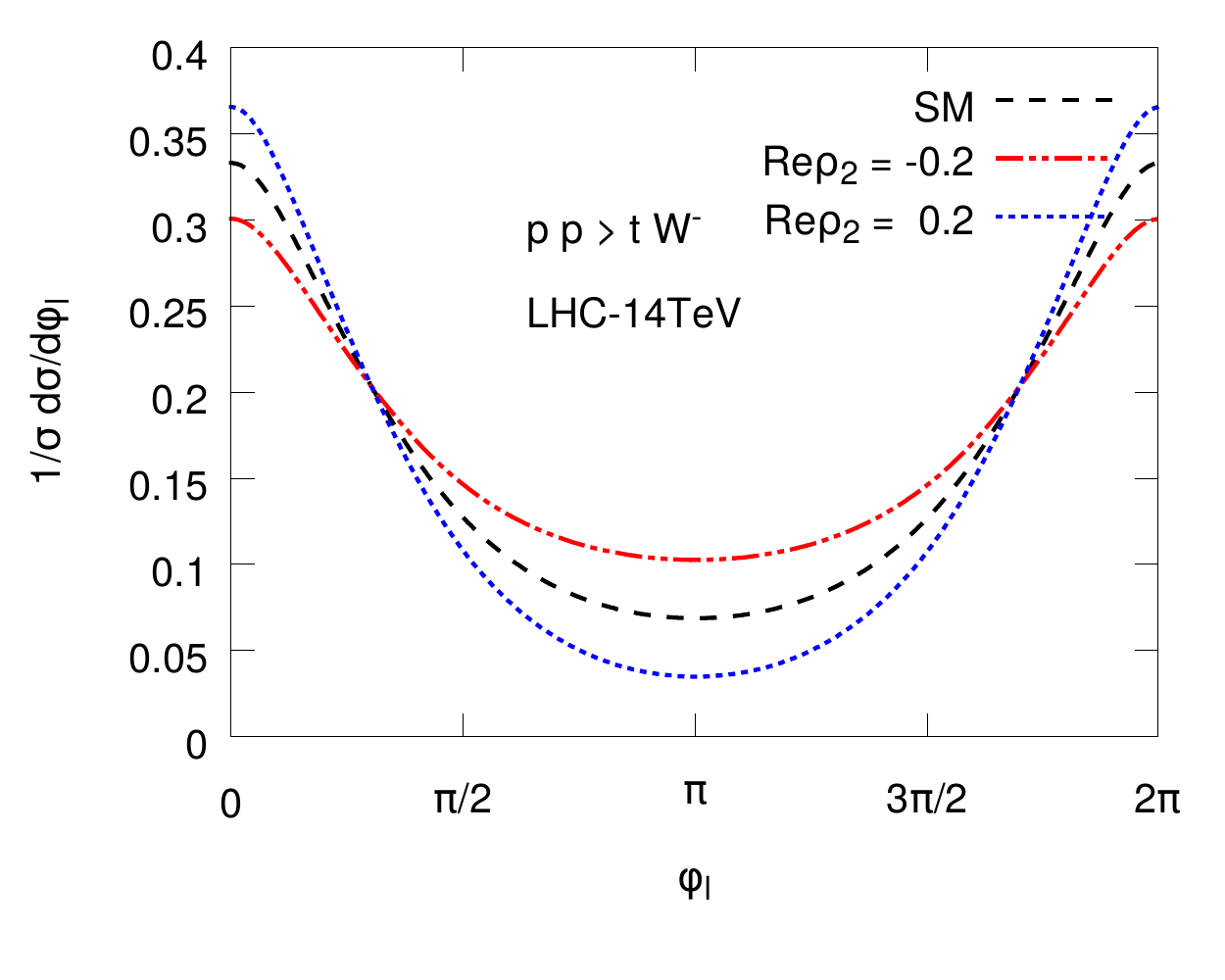} 
\includegraphics[scale=0.65]{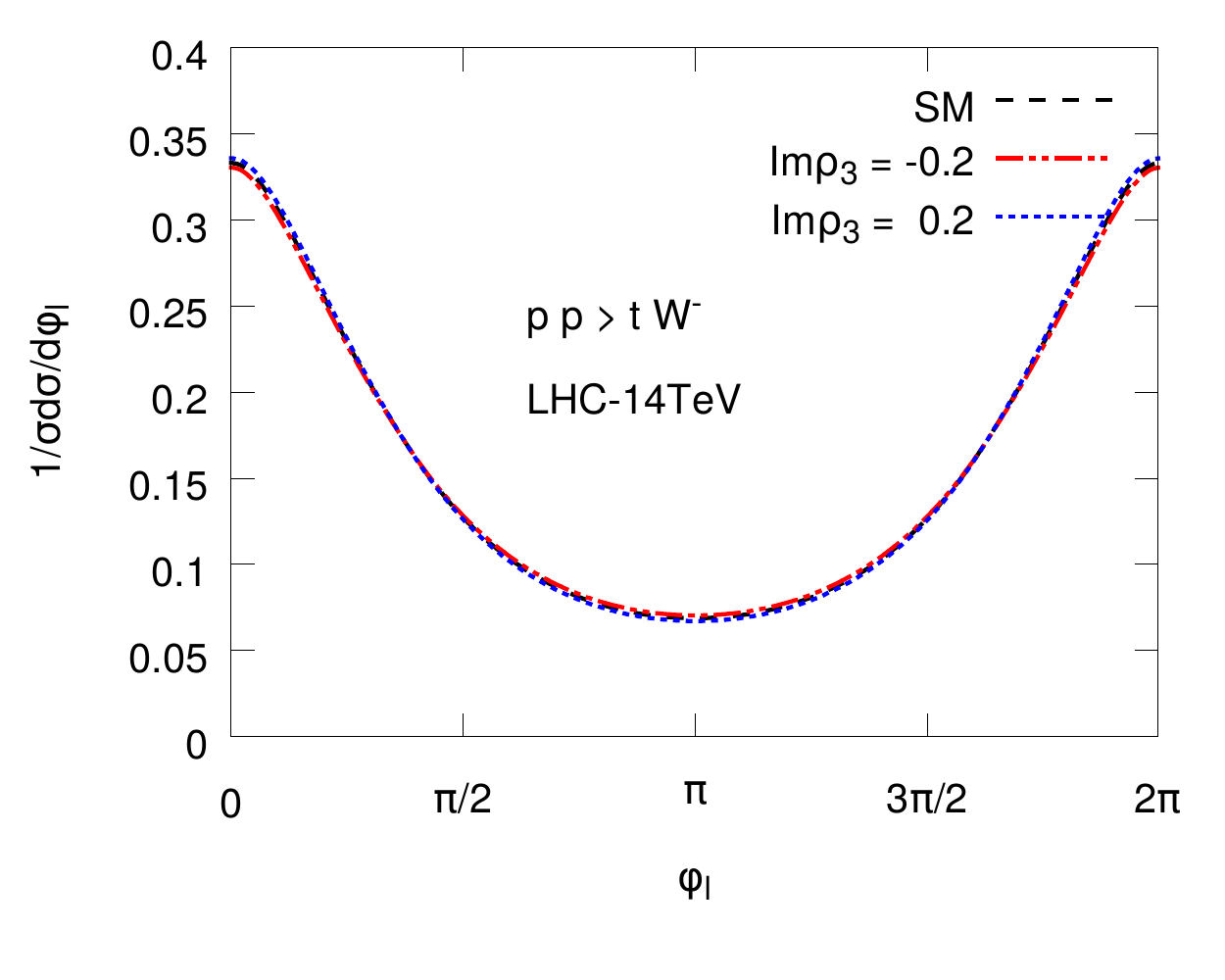} 
\caption{ The normalized azimuthal distribution of the charged lepton in associated-$W$t
single-top production at the LHC14 for anomalous couplings $\f2r$, $\rr$ and $\ri$. The
contributions of anomalous couplings are included up to linear order. Also shown in each case
is the distribution for the SM.} 
\label{dist-azi}
\end{center}
\end{figure}

We next look at the contributions of anomalous couplings to the azimuthal distribution of the 
charged lepton. In Figs. \ref{dist-azi} we show the normalized azimuthal distribution of the 
charged lepton in a linear approximation of the couplings for LHC8 and LHC14 for different
values of $\f2r$, $\rr$ and $\ri$, taken non-zero one at a time. We see that the curves for the
couplings $\f2r$, $\rr$ and $\ri$ peak near $\phi_{\ell}=0$ and $\phi_{\ell}=2\pi$. 
The reason for this is two fold: top polarization and kinematic effect. From Eq. 
(\ref{topdecaywidth}), one finds that the decay lepton prefers to be emitted along the top spin 
direction in the top rest frame, with $\kappa_f=1$. The corresponding distributions in the
parton cm frame are given by Eq. (\ref{angdist}). The rest frame forward (backward) peak
corresponds to a peak for $\cos\theta_{t\ell}=\pm 1$. This is the effect from polarization. 
The kinematic effect is from the $(1-\beta_t\cos\theta_{t\ell})^3$ factor in the denominator
of Eqs. (\ref{angmat1}) and (\ref{angmat2}), which gives rise to peaking for large 
$\cos\theta_{t\ell}$. Also these anomalous couplings, which include momentum dependence, 
further give rise to enhancement or suppression in the top-boost depending on the sign of the
couplings. Thus, for these couplings, there is enhancement or suppression of the peak in the
azimuthal distribution near $\phi_{\ell}=0$ and $\phi_{\ell}=2\pi$. 

Since the couplings $\f2r$ and $\rr$ contribute significantly at the linear order to the
production density matrices $\rho(\lambda,\lambda^\prime)$, the azimuthal distributions for 
$\f2r$ and $\rr$ are quite distinct from the SM at $\phi_{\ell}=0,~2\pi$. On the other hand,
for the coupling $\ri$, the distribution almost overlaps with the SM curve, because its contribution at the linear order to the $\rho(\lambda,\lambda^\prime)$ is quite small. 

\subsection{Azimuthal Asymmetry}

\begin{figure}[h]
\begin{center}
\includegraphics[scale=0.65]{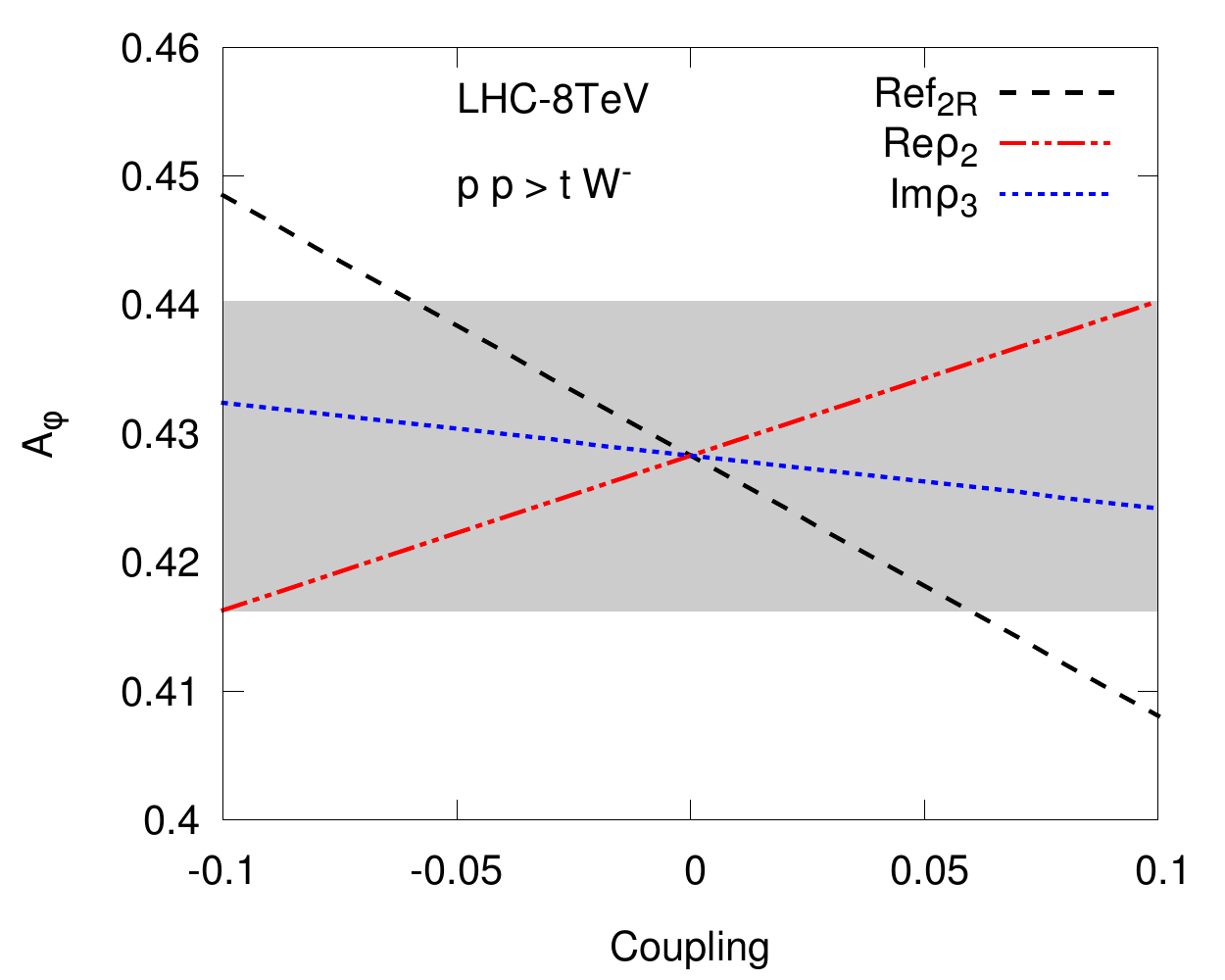} 
\includegraphics[scale=0.65]{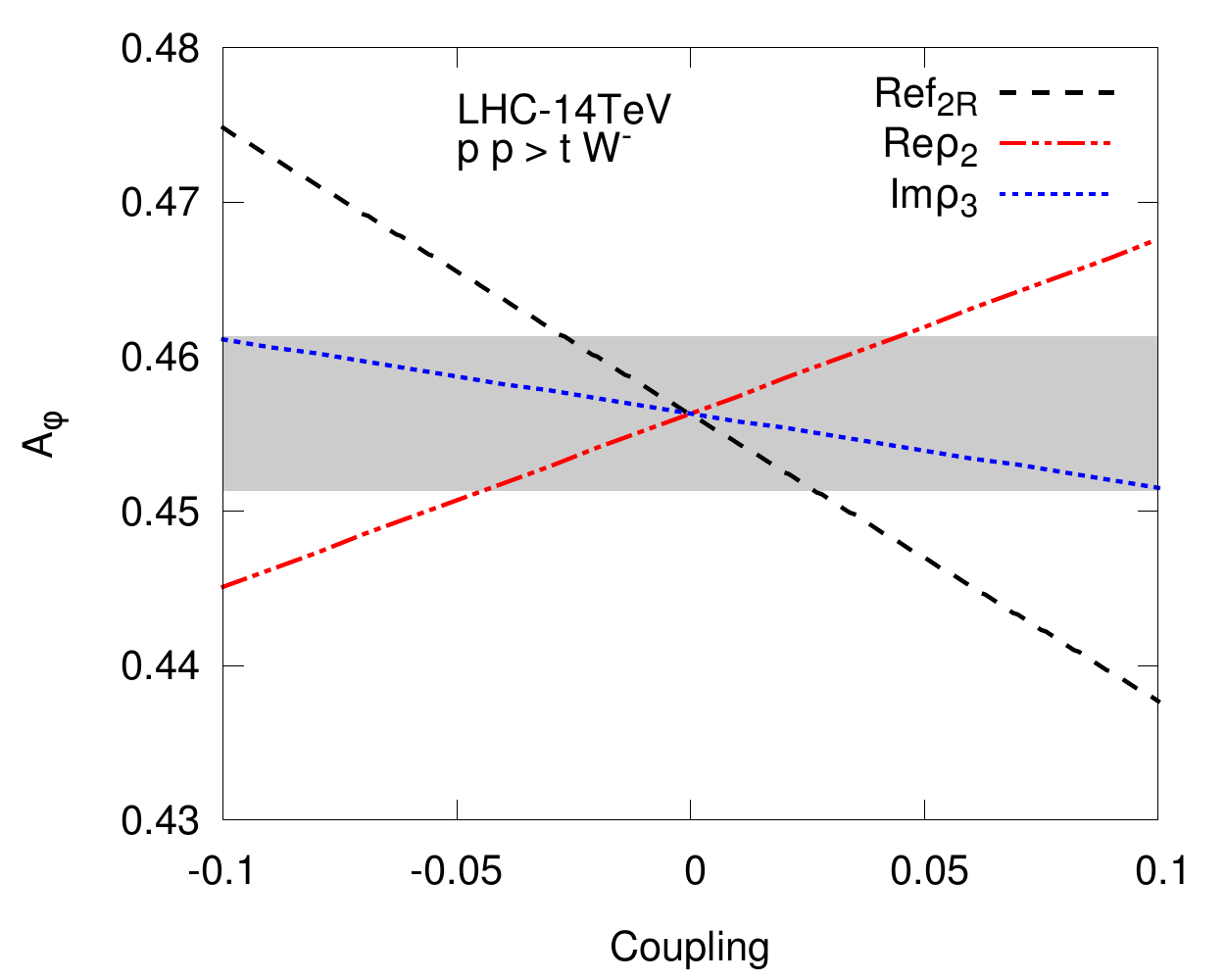} 
\caption{ The azimuthal asymmetry of the charged lepton in associated-$W$t single-top 
production at the LHC8 (left) and LHC14 (right) for different anomalous and $Wtb$ and $ttg$ couplings.} 
\label{phiasym-lin}
\end{center}
\end{figure}

\begin{figure}[h]
\begin{center}
\includegraphics[scale=0.65]{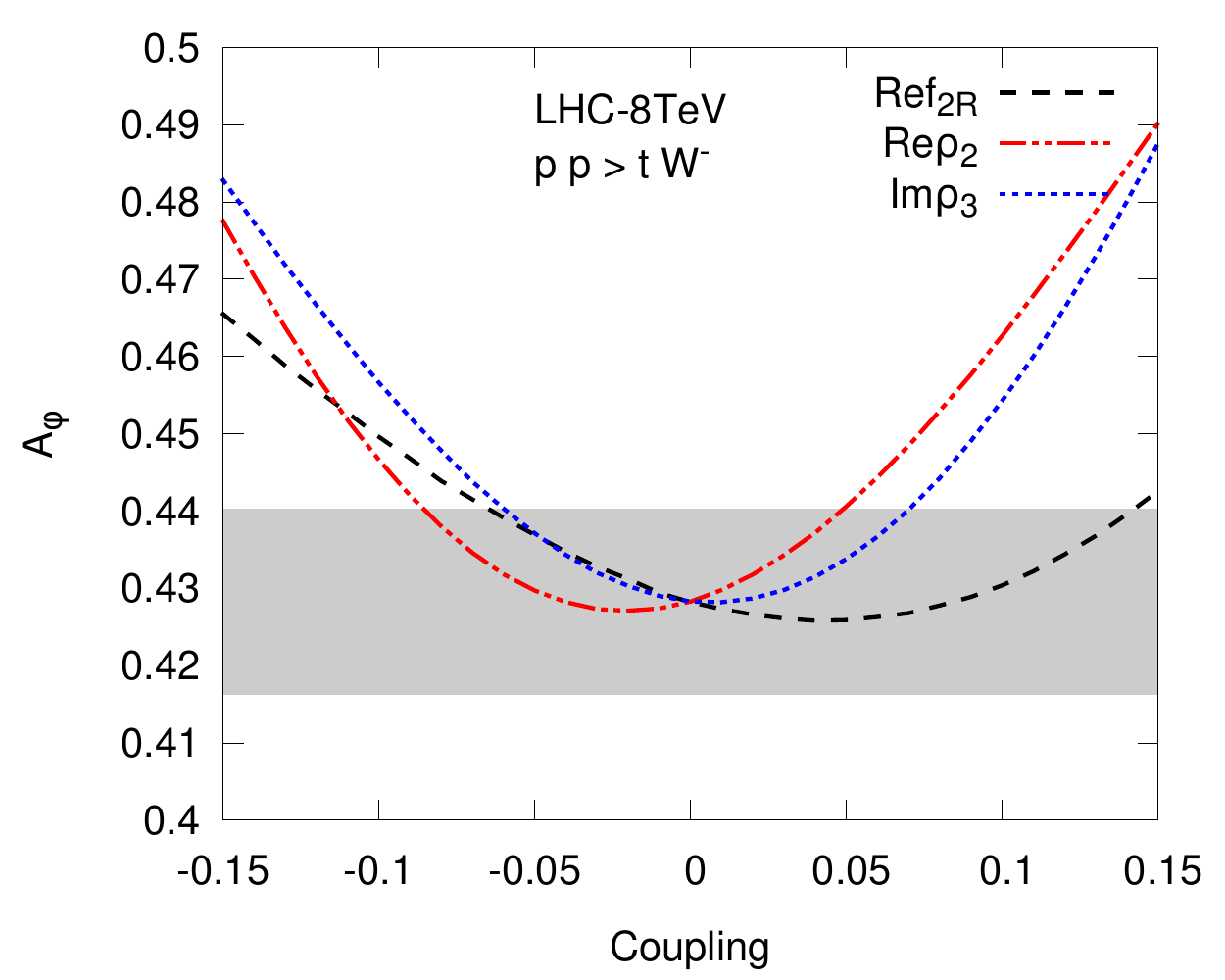} 
\includegraphics[scale=0.65]{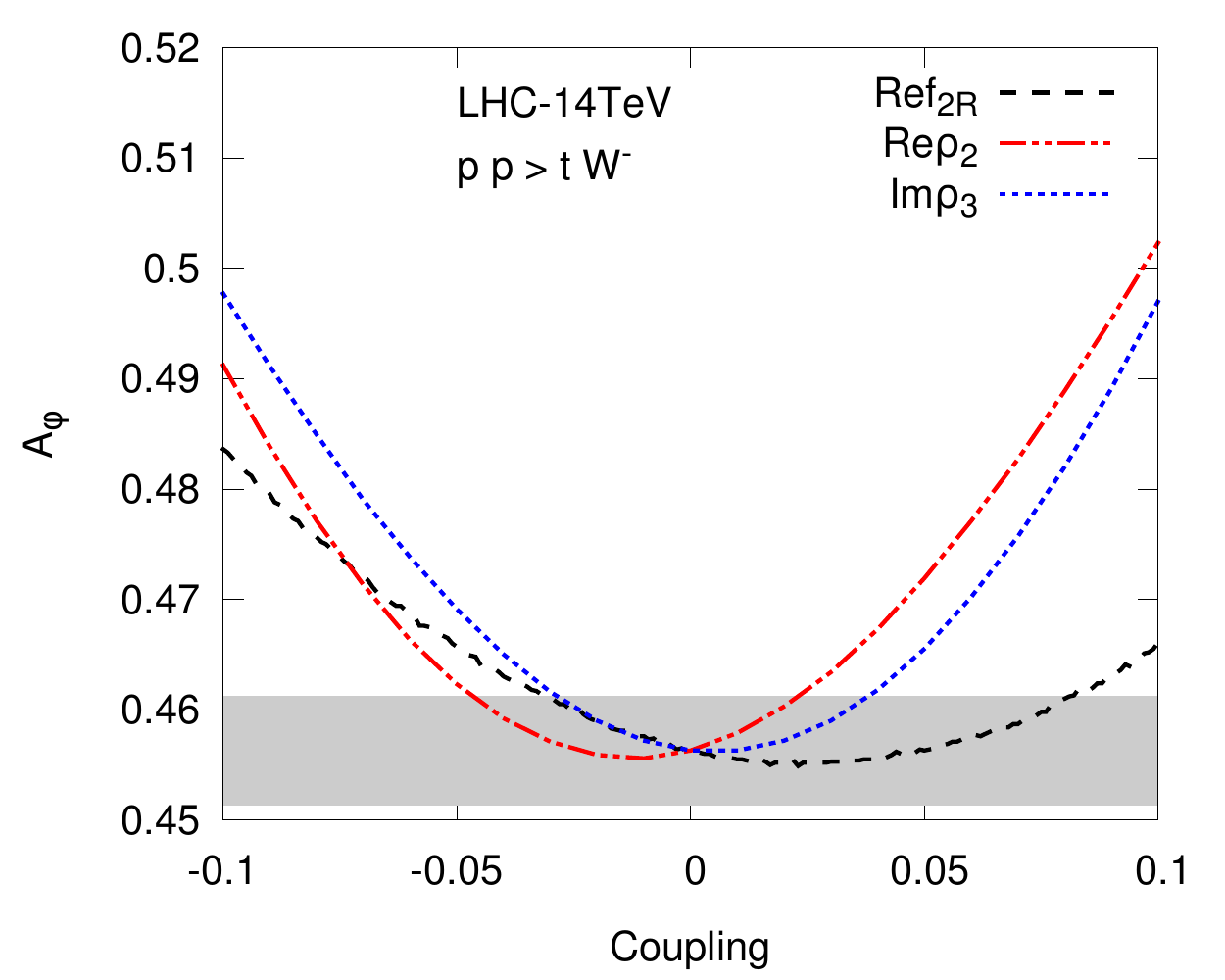} 
\caption{ The azimuthal asymmetry of the charged lepton in associated-$W$t single-top 
production at the LHC8 (left) and LHC14 (right) for different anomalous and $Wtb$ and $ttg$ couplings.} 
\label{phiasym-full}
\end{center}
\end{figure}

As we see from Fig. \ref{dist-azi} that the curves are well 
separated at the peaks for the chosen values of the anomalous $Wtb$ and $ttg$ couplings 
and are also well separated from the curve for the SM. An asymmetry can be defined for
the lepton to quantify these differences in the distributions as
\begin{equation}
 A_{\phi}=\frac{\sigma(\cos \phi_\ell >0)-\sigma(\cos
\phi_\ell<0)}{\sigma(\cos \phi_\ell >0)+\sigma(\cos \phi_\ell<0)},
\label{aziasy}
\end{equation}
where the denominator is the total cross section. This azimuthal asymmetry is in fact the
``left-right asymmetry'' of the charged lepton at the LHC defined with respect to the beam
direction, with the right hemisphere defined as that in which the top momentum lies, and 
the left one being the opposite one. In Fig. \ref{phiasym-lin} we show the plots of 
$A_{\phi}$ at LHC8 and LHC14 as a function of the couplings $\f2r$, $\rr$ and $\ri$ including
their contributions up to the linear order. In Fig. \ref{phiasym-full} plots of 
$A_{\phi}$ are shown including the full contributions of the anomalous couplings at LHC8 and
LHC14. The rapidity and transverse momentum acceptance cuts on the decay 
lepton that we have used to obtain all the distributions and asymmetries are $|\eta|<2.5,\,p_{T}^\ell>20$ GeV.

From Fig. \ref{dist-azi}, we see that the azimuthal distribution of the decay charged lepton 
is more sensitive to $\f2r$, $\rr$ than to $\ri$. Hence, we expect that the azimuthal 
asymmetry we construct in Eq. (\ref{aziasy}) would be a sensitive probe of $\f2r$ and $\rr$.
This fact can indeed be seen from Figs. \ref{phiasym-lin}, where the straight lines for $\rr$ 
and $\f2r$ are steeper than for $\ri$ implying a more significant contribution from the former. 
The reason we get straight lines for individual contributions to the asymmetry is 
that we are working in a linear approximation for the anomalous couplings. 

From Fig. \ref{phiasym-full}, we find that the contributions of the couplings $\rr$ and $\ri$ 
are quite significant for relatively large values ($\sim 0.1$) of the anomalous couplings and the 
linear approximations on these couplings are only valid in the range $[-0.02:0.02]$. Beyond 
this range, quadratic contributions from these couplings also become quite significant. 
$A_\phi$ is more sensitive to negative values of $\f2r$, whereas for $\rr$ it is more sensitive
to positive values of the anomalous coupling. On the other hand, $A_\phi$ is almost symmetric
around $\ri=0$, signifying that there is only a very small contribution at linear order and 
it is equally sensitive to positive and negative values of the coupling. The reason behind is that
the linear contributions from the coupling $\ri$ in the numerator and denominator have 
same sign and this tends to cancel the effect in the asymmetry. The 
grey bands in Figs. \ref{phiasym-lin} and \ref{phiasym-full} denote the statistical 
uncertainty in the measurement of the asymmetry, which has been evaluated using the 
Eq. (\ref{stat-dev}). For LHC8 and LHC14, integrated luminosities of 20 fb$^{-1}$ and 
30 fb$^{-1}$ respectively have been used to evaluate the statistical uncertainties in the measurement of the observables.

\section{\boldmath Sensitivity analysis for anomalous $tbW$ and $ttg$ couplings}\label{sens}
We now study the sensitivities of the observables discussed in the previous sections to the
anomalous $tbW$ and $ttg$ couplings at the LHC, running at two cm of energies viz., 8 TeV and
14 TeV, with integrated luminosities 20 fb$^{-1}$ and 30 fb$^{-1}$, respectively. To obtain 
the 1$\sigma$ limit on the anomalous $tbW$ and $ttg$ couplings from a measurement of an 
observable, we find those values of the couplings for which observable deviates by 1$\sigma$
from its SM value. The statistical uncertainty $\sigma_i$ in the measurement of any generic
asymmetry $\mathcal A_i$ is given by 
\beq\label{stat-dev}
\sigma_i=\sqrt{\frac{1-\left({\mathcal A}^{SM}_i\right)^2}{\mathcal N \epsilon}},
\eeq
where ${\mathcal A}^{SM}_i$ is the asymmetry predicted in the SM, $\mathcal N$ is the total
number of events predicted in the SM and $\epsilon$ is the efficiency of the signal after 
applying all the acceptance and selection cuts to separate the signal from the background. 
We determine the efficiency to be approximately 0.1 by making use of Table 1 of 
\cite{Aad:2012xca} and also Table 1 of \cite{Chatrchyan:2014tua}, where they search for $Wt$ 
using leptonic decays of both the top and $W$. For the searches with hadronic decays of the $W$
and semileptonic decay of the top, which is pertinent 
to our analysis, Ref. \cite{eff} indicates somewhat better efficiency than 0.1 for each channel.
Thus, we are being somewhat conservative as we take $\epsilon=0.1$. 
We apply this to the top polarization, top polar asymmetry and azimuthal asymmetry which
we have discussed. In case of top polarization, the limits are obtained on the
assumption that the polarization can be measured with only leptonic decays and 
thus only the semi-leptonic cross section has been used to calculate the 
statistical uncertainty.

 \begin{table}[h!]
 \begin{center}
 \begin{tabular}{|c|c|c|c|}
\hline
&\multicolumn{3}{c|}{8 TeV}\\
\hline
\hline
 Observable & 	$\f2r$	&  $\rr$ & 	$\ri$	\\
 \hline
 $P_t$ 				&[$-0.030$, $0.032$]		& [$-0.028$, $0.019$] 	&[$-0.038$, $0.065$]\\
 $P_t$ (lin. approx.) 	&[$-0.030$, $0.030$]	& [$-0.022$, $0.022$]	&$[-0.088$, $0.088]$\\
 $A_\phi$ 			&[$-0.060$, $0.140$]	  	& [$-0.080$, $0.050$] 	&[$-0.055$, $0.070$]\\
 $A_{\phi}$ (lin. approx.) &[$-0.065$, $0.065$]	& [$-0.100$, $0.100$]	&[$-0.295$, $0.295$]	\\
\hline
\hline
  \end{tabular}
\caption{Individual limits on anomalous couplings $\f2r$, $\rr$ and $\ri$ which may be 
obtained by the measurement of the observables shown in the first column of the table 
at 8 TeV with integrated luminosities of 20 fb$^{-1}$.}
 \label{lim8TeV}
\end{center}
 \end{table}

\begin{figure}[h!]
\begin{center}
\includegraphics[scale=0.8]{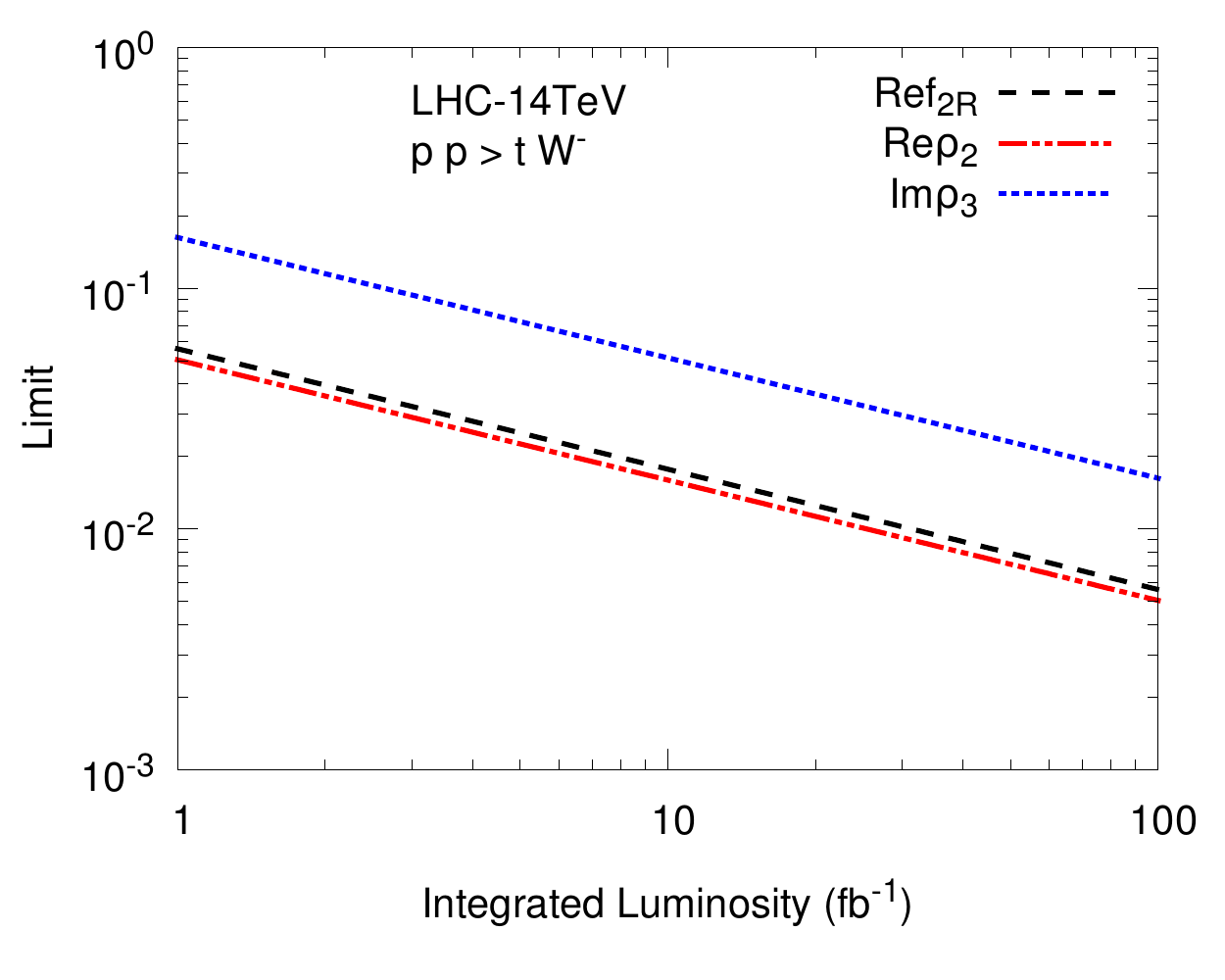} 
\caption{The 1$\sigma$ limit on the anomalous couplings from the measurement of azimuthal
asymmetry $A_\phi$ as a function of integrated luminosity at LHC14.}
\label{lim-lum}
\end{center}
\end{figure}

The 1$\sigma$ limits on $\f2r$, $\rr$ and $\ri$ are given in Table \ref{lim8TeV} and
\ref{lim14TeV} for LHC8 and LHC14, respectively, where we 
assume only one anomalous coupling to be non-zero at a time. We have also assumed measurements
on a $tW^-$ final state. Including the $\bar t W^+$ final state will improve the limits by a
factor of $\sqrt{2}$. In case of the lepton distributions, we take into account only one
leptonic channel. Again, including other leptonic decays of the top would improve the limits
further. The limits corresponding to a linear approximation in the couplings are denoted by
the label ``lin. approx.". Note that quadratic dependence is on $|\rho_2|^2$ and $|
\rho_3|^2$, not on Re$\rho_2$ and Im$\rho_3$. Thus the limits obtained from the quadratic 
expressions assume that Im$\rho_2$ and Re$\rho_3$ are zero. Apart from the 1$\sigma$ limits shown
in Table \ref{lim8TeV}, which correspond to intervals which include zero value of the coupling,
there are other disjoint intervals which could be ruled out if no deviation from the SM is
observed for $P_t$ and $A_{\phi}$. This is apparent from Fig. \ref{polcoup-full}. The additional
allowed intervals for $\f2r$ and $\rr$ from $P_t$ measurement are [0.158, 0.205] and [-0.80,
-0.65] for LHC14, respectively\footnote{[a, b] denotes the allowed values of the coupling $f$ at
the 1$\sigma$ level, satisfying $a<f<b$.}. It is seen that the top polarization, $P_t$, and
azimuthal asymmetry, $A_{\phi}$, of the charged lepton are more sensitive to negative values of
the anomalous couplings $\f2r$ and positive values of $\rr$. In Fig. \ref{lim-lum} we show
projected individual limits (taking only one coupling nonzero at a time) on the anomalous
couplings $\f2r$, $\rr$ and $\ri$, obtained from the measurement of $A_\phi$ at the LHC14 as a
function of integrated luminosity. With 100 fb$^{-1}$ of integrated luminosity, the projected
limits on $\f2r$, $\rr$ and $\ri$ are [-0.006, 0.006], [-0.005, 0.005] and [-0.015, 0.015]
respectively.  

\begin{table}[h!]
\begin{center}
\begin{tabular}{|c|c|c|c|}
\hline
&\multicolumn{3}{c|}{14 TeV}\\
\hline
\hline
 Observable & 	$\f2r$	&  $\rr$ & 	$\ri$	\\
 \hline
 $P_t$ 			&[$-0.010$, $0.010$]	& [$-0.009$, $0.009$] 	&[$-0.020$, $0.035$]\\
 $P_t$ (lin. approx.) 	&[$-0.010$, $0.010$]	& [$-0.009$, $0.009$]	&$[-0.030$, $0.030]$\\
 $A_\phi$ 		&[$-0.031$, $0.081$]	& [$-0.045$, $0.020$] 	&[$-0.030$, $0.040$]\\
 $A_{\phi}$ (lin. approx.) &[$-0.060$, $0.060$]	& [$-0.045$, $0.045$]  	&[$-0.100$, $0.100$]\\
\hline
\hline
  \end{tabular}
\caption{Individual limits on anomalous couplings $\f2r$, $\rr$ and $\ri$ which may be 
obtained by the measurement of the observables shown in the first column of the table 
at 14 TeV with integrated luminosities of 30 fb$^{-1}$.}
 \label{lim14TeV}
\end{center}
 \end{table}

We also obtain simultaneous limits (taking two couplings out of $\f2r$, $\rr$ and $\ri$ 
non-zero simultaneously) on these anomalous couplings that may be obtained by the 
measurements of asymmetries.
For this, we perform a $\chi^2$ analysis to fit all the observables to within $f\sigma$ of
statistical errors in the measurement of the observable. We define the following $\chi^2$
function 
\beq\label{chisq}
\chi^2 = \sum_{i=1}^n\left(\frac{P_i-O_i}{\sigma_i}\right)^2,
\eeq
where the sum runs over the $n$ observables measured and $f$ is the degree of the confidence
interval. The $P_i$'s are the values of the observables obtained by taking two couplings out of 
$\f2r$, $\rr$ and $\ri$ non-zero simultaneously and the $O_i$'s are the values of the observables 
obtained in the SM. The $\sigma_i$'s are the statistical fluctuations in the measurement of the 
observables, given in Eq. (\ref{stat-dev}).

\begin{figure}[h!]
\begin{center}
\includegraphics[scale=0.55]{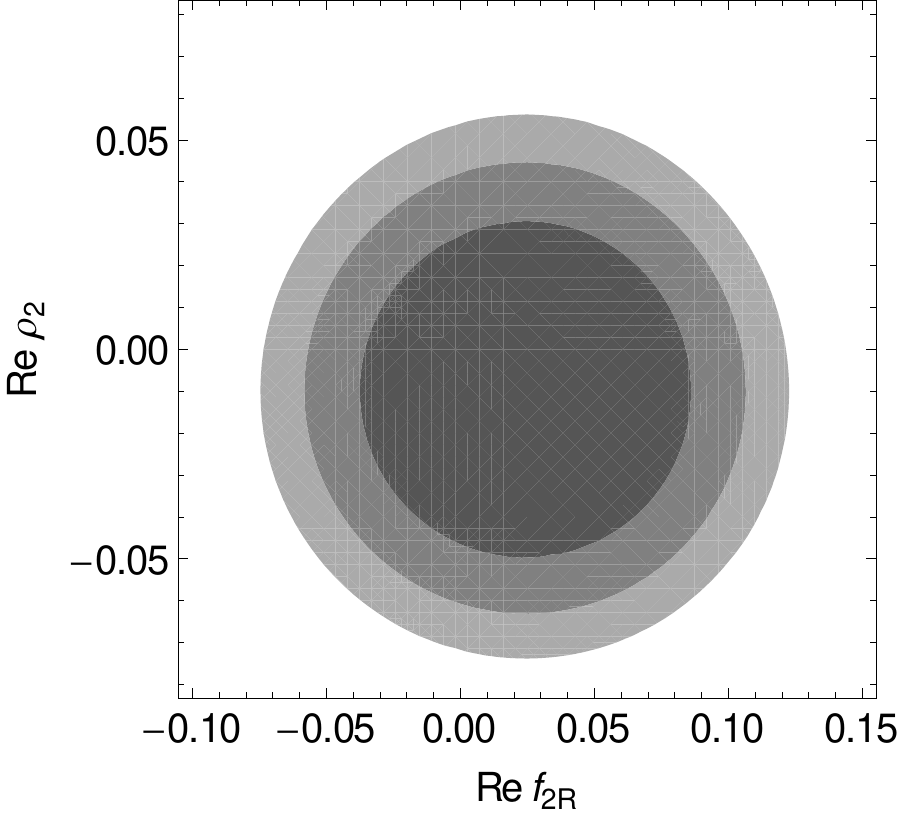} 
\includegraphics[scale=0.55]{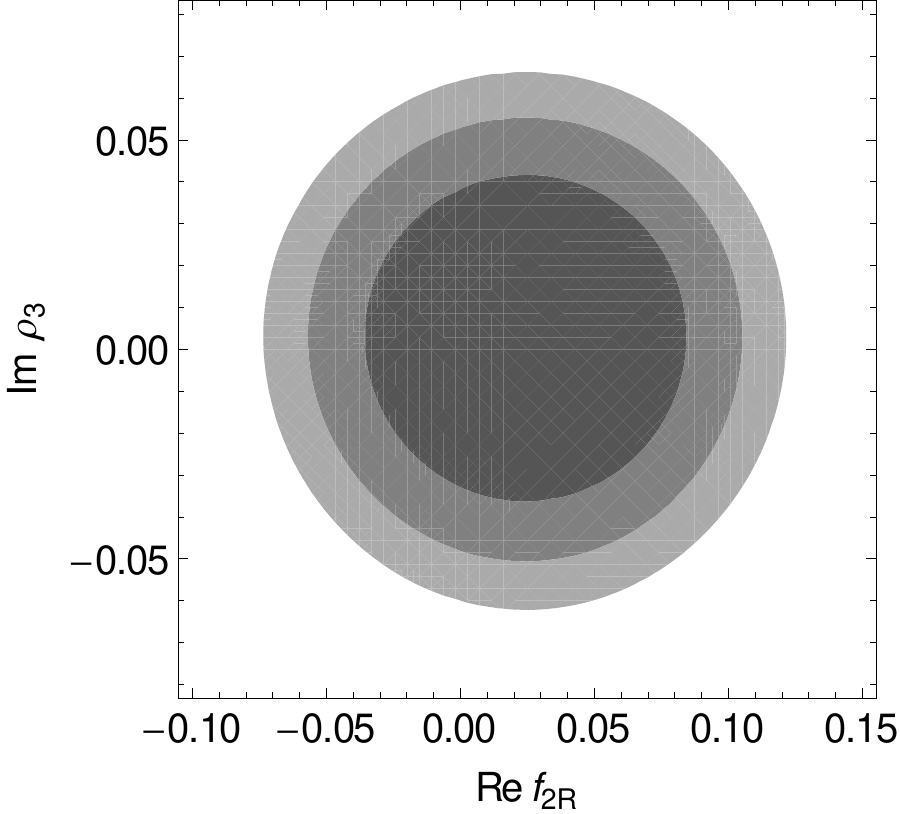} 
\includegraphics[scale=0.55]{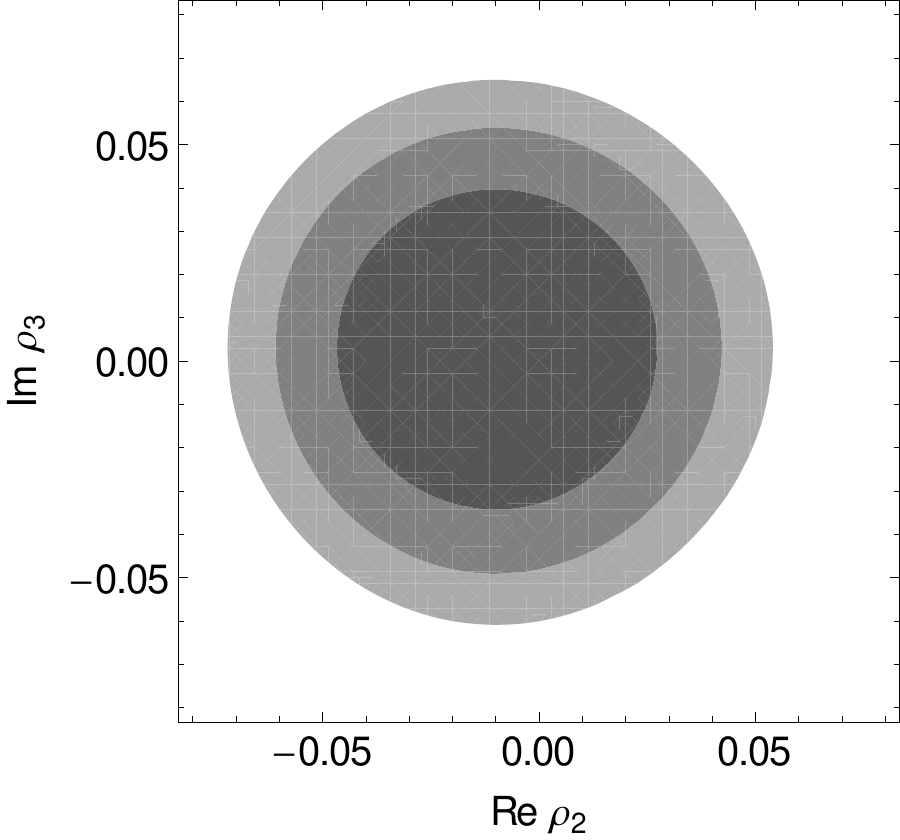} 
\caption{The 1$\sigma$ (central region), 2$\sigma$ (middle region) and 3$\sigma$ (outer
region) CL regions in the $\f2r$-$\rr$ plane (left), $\f2r$-$\ri$ plane (center) and 
$\rr$-$\ri$ plane (right) allowed by the measurement of the azimuthal asymmetry at the 
LHC14. The $\chi^2$ value for the 1$\sigma$, 2$\sigma$ and 3$\sigma$ CL intervals are 2.30, 
6.18 and 11.83 respectively, for two parameter fit.} 
\label{lim-all}
\end{center}
\end{figure}

In Fig. \ref{lim-all}, we show the 1$\sigma$, 2$\sigma$ and 3$\sigma$
regions in $\f2r-\rr$ plane, $\f2r-\ri$ plane and $\rr-\ri$ plane allowed by the measurement of
the asymmetry $A_\phi$. From the plots shown in Fig. \ref{lim-all}, we find that the strongest
simultaneous limits are [$-0.03$, 0.08] on $\f2r$, [$-0.05$, 0.02] on $\rr$  and [$-0.03$, 0.03]
on $\ri$, at the $1\sigma$ level.

We now compare our results with those of other relevant works on the determination of anomalous
$ttg$ couplings at the LHC. Ref. \cite{Fabbrichesi:2013bca} studies the top quark compositeness 
and put a stringent limit of $|0.01|$ on top-CMDM coupling using top-pair production cross section and spin correlations at LHC7 and LHC8. In Ref. 
\cite{Biswal:2012dr}, the authors used top polarization asymmetry and azimuthal asymmetry in top
pair production to probe $\rr$ and $\ri$ at the Tevatron and LHC. Their conclusion was that top
polarization $P_t$ is only sensitive to $\ri$ and not $\rr$. On the other hand, in the $Wt$-mode
of single-top production, we find opposite results. Here $P_t$ is more sensitive to $\rr$ than to
$\ri$. We also find that the limits which we obtained on anomalous couplings are about an order of
magnitude smaller than those obtained in Ref \cite{Biswal:2012dr} for top pair production from the
observables considered in this work. In Ref. \cite{Rizzo:1995uv} Rizzo studied anomalous $ttg$
couplings in single-top production at the Tevatron and at the LHC and concluded that the limits
from this channel are about one order of magnitude smaller than those from the pair production
processes. 

\section{Conclusions}\label{conclusions}

We have investigated the sensitivity of the LHC8 and LHC14 to the anomalous $ttg$ couplings in
$Wt$ mode of single-top production followed by semileptonic decay of the top. We derived
analytical expressions for the spin density matrix for single-top-quark production, including the
contributions of both real and imaginary parts of the anomalous $ttg$ couplings. We find that only
$\rr$ and $\ri$ give significant contributions to the spin density matrix at linear order. It may
be noted that Im$\rho_2$ and Re$\rho_3$ do not appear in the observables we consider. This may be
understood from the CPT theorem which implies that our observables, being even under naive time
reversal T, can only get a non vanishing contribution from dispersive parts of CP-even form
factors (in this case $\rr$) and from absorptive parts (in this case $\ri$) of CP-odd form
factors. 

Since top polarization can be measured only through the differential distribution of its decay
products, we also study the angular distributions of the charged lepton coming from the decay of
the top. We mainly focus on charged-lepton distribution for three reasons : a) charged-lepton
momenta are very accurately measured at the LHC, b) they have  the best spin analyzing power and
c) their angular distributions have been shown to be independent of any new physics in top decay.
We find that the polar-angle distribution is not very sensitive to the anomalous couplings. On the
other hand, the normalized azimuthal distribution is found to be sensitive to the anomalous
couplings. The azimuthal distribution peaks close to $\phi=0$ and $\phi = 2\pi$ and the values at
the peaks are quite sensitive to the magnitude and the sign of the anomalous couplings. In order
to quantify this difference and to be statistically more sensitive, we construct an integrated
azimuthal asymmetry from the azimuthal distribution of charged lepton. 

Our observables, in this work, are constructed from one of the two conjugate single-top 
processes possible, viz., $tW^-$ in the final state. Under a CP transformation, since helicity
changes sign, a positive helicity top quark transforms to a negative helicity top anti-quark. If
CP is conserved, the spin density matrix, $\bar \sigma$, for the top anti-quark in the 
$\bar t W^+$ production process would be obtained from that of the top quark in the $t W^-$
production process by reversing the sign of helicities i.e., $\bar\sigma^{\pm,\pm}=\sigma^{\mp,
\mp}$ and $\sigma^{\pm,\mp}=\sigma^{\mp,\pm}$. However, since $\ri$ is CP odd, the contribution of
$\ri$ in the $\bar\sigma$ would change sign. Also, because of CPT transformation properties, the
signs of $\mathrm{Im}\rho_2$ and $\mathrm{Re}\rho_3$ in the $\bar \sigma^{\pm\mp}$ also change.
If we combine observables from $tW^-$ production as
well as $\bar tW^+$ production, it is possible to construct observables with definite CP 
properties enabling us to separate contributions from $\rho_2$ and $\rho_3$, which have opposite
CP transformation properties. The simplest of these would be the sum (difference) of the $tW^-$
and $ tW^+$ production cross sections which would be CP even (odd) and the sum (difference) of $t$
and $\bar t$ polarizations in the two processes, which would be CP odd (even). With the
availability in future of large event samples in the high-luminosity version of the experiment, it
would be possible to separately constrain CP-even and CP-odd couplings. 

Our proposal provides an alternative other than top-pair production to look for top CMDM and CEDM
in single-top production. The $Wt$ mode is also significant in the probe of these couplings
because it does not get any contributions from other new physics like new resonances, exotic
quarks or scalars which could contaminate other single-top production modes and $t\bar t$-
production process.  

We also note that the production cross section at 13 TeV is about 15\% lower than the 14 TeV cross section, 
while the asymmetry and top polarization are not very sensitive to the cm energy of the LHC. 
The latter change by less than a percent. Thus, the sensitivities would only be affected by the 15\% 
reduction in the total number of events at 13 TeV relative to 14 TeV LHC for the same amount of 
integrated luminosity. This would amount to a reduction in sensitivity to the couplings of about 7\%.

In conclusion, we have shown that top polarization, and subsequent decay-lepton distributions can
be used to obtain fairly stringent limits on chromomagnetic and chromoelectric top couplings from
the existing 8 TeV run of the LHC. The limits could be improved by the future runs of the LHC at
14 TeV.

Our results would be somewhat worsened by the inclusion of realistic detection efficiencies for
the $b$ jet and for the detection of the $W$. 
On the other hand, inclusion of the $\bar tW^+$ final state, as well as additional leptonic
channels in top decay would contribute to improving on our estimates of the limits. A more
complete analysis including detector simulation would be worthwhile to carry out.

\section*{Acknowledgements}
SDR acknowledges support from the Department of
Science and Technology, India, under the J.C. Bose National
Fellowship programme, Grant No. SR/SB/JCB-42/2009. This work was supported by the Australian Research Council through the ARC Centre of Excellence in Particle Physics at the Terascale (CE110001004) and grant FL0992247

\section*{Appendix: Spin-density matrix elements of top quark in tW production including
anomalous top-gluon couplings}
In the appendix, we give the spin density matrix elements for the $tW$ single-top production
process. We include the contributions of top CMDM and CEDM up to the linear order. The 
diagonal elements of the spin density matrix can be written as 
\begin{equation}\label{rhocomponents}
\rho^{\pm\pm}=\rho^{\pm\pm}_{s,SM}+\rho^{\pm\pm}_{t,SM}+\rho^{\pm\pm}_{st,SM}+
\rho^{\pm\pm}_{t,\rr}+\rho^{\pm\pm}_{st,\rr}+\rho^{\pm\pm}_{t,\ri}+\rho^{\pm\pm}_{st,\ri}
\end{equation}
where $s$, $t$ and $st$ in the subscript denote the contribution from $s$-channel, $t$-channel
and interference between $s$ and $t$ channels respectively.

\begin{eqnarray}
\rho_{s,SM}^{\pm\pm}&=& \frac{g^2 \ g_s^2}{24 \ \shat} \ \frac{1}{m_W^2}
\Big[(\that + \uhat + m_W^2) \ (\that-m_W^2) +\shat \ m_W^2\nonumber\\
&\pm&  2\mt \left \{\ \pgn3\left(\that - 2m_W^2\right) 
+ \pbn3\left(\that - m_W^2\right)\right\}\Big]
\end{eqnarray}
\begin{eqnarray}
\rho_{t,SM}^{\pm\pm}&=& \frac{g^2 \ g_s^2}{24 \ (\that -\mt^2)^2} \ \frac{1}{\mt^2 \ m_W^2}
\Big[\mt^2(\uhat - m_W^2)(\mt^2 -4m_W^2-4\uhat)
+\uhat^2(\mt^2 - \that) -\shat \ \uhat \ \mt^2\nonumber\\
&+&2\shat \ m_W^2(\mt^2 - \uhat)\pm 2\mt \ \pgn3(\mt^2-2m_W^2)(m_W^2-\uhat)\nonumber\\
&\pm & 2\mt \ \pbn3 (2m_W^2 - \uhat)(\mt^2+\uhat)\Big]\\
&&\nonumber\\
\rho^{\pm\pm}_{st,SM}&=&\nonumber\frac{g^2 \ g_s^2}{24 \ (\that -\mt^2)} \ 
\frac{2}{ \ m_W^2}
\Big[\shat \Big\{(\mt^2 - \that)(\mt^2 +2m_W^2)+\uhat \ \mt^2+\mt^4
+\that\uhat)\Big\}\\\nonumber
&+&(-2\mt^2+\uhat+\that)(\mt^2-\that)(\mt^2+2m_W^2)
\pm\mt \ \pgn3\Big\{(-2\mt^2+2m_W^2+\uhat)(\mt^2-\that)
+2\mt^2\shat\Big\}\\&\pm&\mt\pbn3\Big\{-2(\uhat-m_W^2)(\mt^2-2m_W^2)
+(2m_W^2 - \uhat)(\mt^2 - \uhat)+2\shat\uhat\Big\}\Big]
\end{eqnarray}

\begin{eqnarray}
\rho_{t,\rr}^{\pm\pm}&=& \frac{g^2 \ g_s^2}{24 \ (\that -\mt^2)^2} \ 
\frac{\rr}{\mt^2 \ m_W^2}\Big[3\mt^2 \ 
\left[(\mt^2-\uhat)^2(\mt^2 - \that) 
+\shat(\mt^2 - \uhat)(2m_W^2-\mt)\right]\nonumber\\
&\pm& \mt \ 
\left\{\pgn3\left[(2 m_W^2 -2\mt^2-\uhat)(\mt^2-\uhat)(\mt^2-\that)\right.\right.\\
&+&\left.\left.2\shat \ \mt^2(\mt^2 -4 m_W^2+\uhat)\right]
\pm \pbn3 \left[(\uhat-2 m_W^2)(\mt^2 - \uhat)^2\right]\right\}\Big]\nonumber\\
&&\nonumber\\
\rho^{\pm\pm}_{st,\rr}&=&\frac{g^2 \ g_s^2}{24 \ (\that -\mt^2)} \ \frac{\rr}{\mt \ m_W^2}
\Big[\mt\shat\Big\{(\that-m_W^2)(\mt^2-2m_W^2)-2m_W^2\mt^2+3\uhat\that-\mt^2\that\Big\}
\nonumber\\\nonumber 
&\pm& 2 \left\{(\mt^2-\that)(\mt^2-2m_W^2)-2m_W^2(\mt^2-\uhat)\right\}
\Big\{(\mt^2-\that)\pgn3-(\mt^2-\uhat)\pbn3\Big\}\\\nonumber
&\pm &\shat\left\{(\mt^2-\uhat)\pbn3-(\mt^2-\that)\pgn3\right\}(\mt^2-2m_W^2)
\pm\shat \ \pgn3(3\mt^2m_W^2+\that\uhat-\mt^2\shat)\\
&\pm&\shat \ \pbn3\{\mt^4-u^2-4\mt^2(2m_W^2-\uhat)\}\Big]
\end{eqnarray}
\begin{eqnarray}
\rho_{t,\ri}^{\pm\pm}&=& \frac{g^2 \ g_s^2}{24 \ (\that -\mt^2)^2} \ \frac{\ri}{\mt^2 \ m_W^2}
\Big[\mt^2\left\{(\mt^2 - \uhat)\left[(\mt^2-\that)
(4m_W^2-3\mt^2 +\uhat)+\shat (\ \mt^2 - 2m_W^2 - 2\uhat)\right]\right.\nonumber\\
&+&\left.2\shat \ \mt^2(\mt^2-2 m_W^2)\right\}\pm \mt \pgn3\left\{-2\shat \ \mt^2
(\mt^2 -3\uhat -4m_W^2) +(\mt^2 -\that)\right.\\
&\times &\left. [(2\mt^2-6m_W^2-\uhat)(\mt^2-\uhat)-4\uhat(2m_W^2+\mt^2)]\right\}
\pm \mt\pbn3(\uhat-2m_W^2)(\mt^2 -\uhat)^2\Big]\nonumber\\
&&\nonumber\\
\rho^{\pm\pm}_{st,\ri}&=&\nonumber\frac{g^2 \ g_s^2}{24 \ (\that -\mt^2)} \ 
\frac{\ri}{\mt \ m_W^2}
\Big[\mt\shat\Big\{m_W^2(\mt^2+2m_W^2)-6m_W^2\mt^2+2\uhat\mt^2+(2m_W^2 - \uhat)\that
-2\shat\uhat \Big\}\\\nonumber
&+&2 \ \pgn3\Big\{(\that-\mt^2)(\mt^2+2m_W^2)+2 \ m_W^2(\uhat-\mt^2)
+\shat(2\mt^2+m_W^2)\Big\}(\mt^2-\that)\\\nonumber
&+&\shat \ \pgn3(\{3\uhat -2\shat\} \ \mt^2+\that\uhat)
+\pbn3\Big\{(-2\mt^2 - 2m_W^2 + \uhat)\shat+2(\mt^2-\that)(\mt^2+2m_W^2)\\
&+&4m_W^2(\mt^2-\uhat)\Big\}(\mt^2-\uhat)\Big]\\\nonumber
\end{eqnarray}
where $\shat$, $\that$ and $\uhat$ are the Mandelstam variables in the parton 
cm frame, 
$\pbn3=p_b\cdot n_3$ and $\pgn3=p_g\cdot n_3$. 
$n_3$ is the longitudinal spin vector of the top quark, whose components
are given by 
\begin{equation}
n_3 \equiv \frac{E_t}{m_t}\left(\beta_t,  \sin\theta_t, 0, \cos\theta_t \right).
\end{equation}.

The off-diagonal elements with helicity combination $\pm\mp$ 
corresponding to each of the terms on the
right-hand side of Eq. (\ref{rhocomponents}) are given by $n_3$ dependent terms of 
the diagonal elements for the corresponding term in Eqs. (17)-(23), with the vector
$n_3$ replaced by $\frac{1}{2} (n_1 \mp i \ n_2)$, where
\begin{equation}
n_1 \equiv ( 0, \cos\theta_t, 0, -\sin\theta_t )
\end{equation}
and
\begin{equation}
n_2 \equiv ( 0, 0, 1, 0).
\end{equation}

Moreover there are additional contributions to the spin density matrix elements for helicity
combination $\pm\mp$ which come from couplings $\mathrm{Im \rho_2}$ and $\mathrm{Re \rho_3}$ 
and are given by

\begin{eqnarray}
\rho^{\pm\mp}_{t,\mathrm{Im \rho_2}}&=&\frac{g^2 \ g_s^2}{24 \ (\that -\mt^2)^2} \ 
\frac{\mathrm{Im \rho_2}}{\mt \ m_W^2} \epsilon_{\mu\nu\rho\sigma}
p_t^\mu p_b^\nu p_g^\rho (n_1^\sigma\mp i \ n_2^\sigma)
\Big[(2m_W^2-\uhat)(\mt^2-u)\Big]\\\nonumber
&&\\
\rho^{\pm\mp}_{t,\mathrm{Re \rho_3}}&=&\frac{g^2 \ g_s^2}{24 \ (\that -\mt^2)^2} \ 
\frac{\mathrm{Re \rho_3}}{\mt \ m_W^2} \epsilon_{\mu\nu\rho\sigma}
p_t^\mu p_b^\nu p_g^\rho (n_1^\sigma\mp i \ n_2^\sigma)
\Big[(2m_W^2-\uhat)(3\mt^2+u)\Big]\\\nonumber
\end{eqnarray}

\begin{eqnarray}
\rho^{\pm\mp}_{st,\mathrm{Im \rho_2}}&=&\frac{g^2 \ g_s^2}{24 \ (\that -\mt^2)} \ 
\frac{\mathrm{Im \rho_2}}{\mt \ m_W^2} \epsilon_{\mu\nu\rho\sigma}
p_t^\mu p_b^\nu p_g^\rho (n_1^\sigma\mp i \ n_2^\sigma)
\Big[(\mt^2 + 2m_W^2)(-2\mt^2 + 2\that +\shat)\nonumber\\
&+&(\shat - 4m_W^2)(\mt^2-\uhat)]\\\nonumber
&&\\
\rho^{\pm\mp}_{st,\mathrm{Re \rho_3}}&=&\frac{g^2 \ g_s^2}{24 \ (\that -\mt^2)} \ 
\frac{\mathrm{Re \rho_3}}{\mt \ m_W^2} \epsilon_{\mu\nu\rho\sigma}
p_t^\mu p_b^\nu p_g^\rho (n_1^\sigma\mp i \ n_2^\sigma)
\Big[(\mt^2 - 2m_W^2)(-2\mt^2+2\that+\shat)\nonumber\\
&+&4m_W^2(\mt^2-\uhat)+(\mt^2+\uhat)\shat]\\\nonumber
\end{eqnarray}


\begin{thebibliography}{10}

\bibitem{top:mass} 
  [Tevatron Electroweak Working Group and CDF and D0 Collaborations],
  arXiv:1107.5255 [hep-ex].

\bibitem{atlas-spin} 
  G.~Aad {\it et al.}  [ATLAS Collaboration],
  Phys.\ Rev.\ Lett.\  {\bf 108}, 212001 (2012)
  [arXiv:1203.4081 [hep-ex]].

\bibitem{cms-spin}
CMS Collaoration, CMS Physics Analysis Summary, CMS-PAS-TOP-12-004.


\bibitem{LHC-toppol}
ATLAS Collaboration, Atlas note, ATLAS-CONF-2012-133;
CMS Collaboration, CMS Physics Analysis Summary, CMS-PAS-TOP-12-016.

\bibitem{Bernreuther:2008ju} 
  W.~Bernreuther,
  J.\ Phys.\ G {\bf 35}, 083001 (2008)
  [arXiv:0805.1333 [hep-ph]].

\bibitem{Bernreuther:2015wqa} 
  W.~Bernreuther and P.~Uwer,
  Nucl.\ Part.\ Phys.\ Proc.\  {\bf 261-262}, 414.

\bibitem{Bernreuther:2013aga} 
  W.~Bernreuther and Z.~G.~Si,
  Phys.\ Lett.\ B {\bf 725}, 115 (2013)
  [Phys.\ Lett.\ B {\bf 744}, 413 (2015)]
  [arXiv:1305.2066 [hep-ph]].

\bibitem{Godbole:2010kr} 
  R.~M.~Godbole, K.~Rao, S.~D.~Rindani and R.~K.~Singh,
  JHEP {\bf 1011}, 144 (2010)
  [arXiv:1010.1458 [hep-ph]].

\bibitem{Huitu:2010ad} 
  K.~Huitu, S.~Kumar Rai, K.~Rao, S.~D.~Rindani and P.~Sharma,
  JHEP {\bf 1104}, 026 (2011)
  [arXiv:1012.0527 [hep-ph]].

\bibitem{Godbole:2011vw} 
  R.~M.~Godbole, L.~Hartgring, I.~Niessen and C.~D.~White,
  JHEP {\bf 1201}, 011 (2012)
  [arXiv:1111.0759 [hep-ph]].

\bibitem{Rindani:2011pk} 
  S.~D.~Rindani and P.~Sharma,
  JHEP {\bf 1111}, 082 (2011)
  [arXiv:1107.2597 [hep-ph]].
  
\bibitem{Prasath:2014mfa} 
  A.~Prasath, R.~M.~Godbole and S.~D.~Rindani,
  arXiv:1405.1264 [hep-ph];
  R.~M.~Godbole, S.~D.~Rindani and R.~K.~Singh,
  Phys.\ Rev.\ D {\bf 67}, 095009 (2003)
  [Erratum-ibid.\ D {\bf 71}, 039902 (2005)]
  [hep-ph/0211136].

\bibitem{Rindani:2011gt} 
  S.~D.~Rindani and P.~Sharma,
  Phys.\ Lett.\ B {\bf 712}, 413 (2012)
  [arXiv:1108.4165 [hep-ph]].

\bibitem{Biswal:2012dr} 
  S.~S.~Biswal, S.~D.~Rindani and P.~Sharma,
  Phys.\ Rev.\ D {\bf 88}, 074018 (2013)
  [arXiv:1211.4075 [hep-ph]].
 
\bibitem{Rindani:2013mqa} 
  S.~D.~Rindani, R.~Santos and P.~Sharma,
  JHEP {\bf 1311}, 188 (2013)
  [arXiv:1307.1158].
  
\bibitem{top:fba} 
  D.~Choudhury, R.~M.~Godbole, S.~D.~Rindani and P.~Saha,
  Phys.\ Rev.\ D {\bf 84}, 014023 (2011)
  [arXiv:1012.4750 [hep-ph]];
  D.~-W.~Jung, P.~Ko and J.~S.~Lee,
  Phys.\ Lett.\ B {\bf 701}, 248 (2011)
  [arXiv:1011.5976 [hep-ph]];
  R.~M.~Godbole, G.~Mendiratta and S.~Rindani,
  arXiv:1506.07486 [hep-ph];
  J.~Cao, K.~Hikasa, L.~Wang, L.~Wu and J.~M.~Yang,
  Phys.\ Rev.\ D {\bf 85}, 014025 (2012)
  [arXiv:1109.6543 [hep-ph]].



\bibitem{Heinson:1996zm}
  A.~Heinson, A.~S.~Belyaev, E.~E.~Boos,
  Phys.\ Rev.\  {\bf D56}, 3114-3128 (1997) 
  [hep-ph/9612424].

\bibitem{Stelzer:1998ni}
  T.~Stelzer, Z.~Sullivan and S.~Willenbrock,
  Phys.\ Rev.\  D {\bf 58}, 094021 (1998)
  [arXiv:hep-ph/9807340].

\bibitem{Belyaev:1998dn}
  A.~S.~Belyaev, E.~E.~Boos and L.~V.~Dudko,
  Phys.\ Rev.\  D {\bf 59}, 075001 (1999)
  [arXiv:hep-ph/9806332].

\bibitem{Boos:1999dd}
  E.~Boos, L.~Dudko and T.~Ohl,
  Eur.\ Phys.\ J.\  C {\bf 11}, 473 (1999)
  [arXiv:hep-ph/9903215].

\bibitem{Tait:1999cf}
  T.~M.~P.~Tait,
  Phys.\ Rev.\  D {\bf 61}, 034001 (2000)
  [arXiv:hep-ph/9909352].

\bibitem{Espriu:2001vj}
  D.~Espriu and J.~Manzano,
  Phys.\ Rev.\  D {\bf 65}, 073005 (2002)
  [arXiv:hep-ph/0107112].

\bibitem{Espriu:2002wx}
  D.~Espriu and J.~Manzano,
  Phys.\ Rev.\  D {\bf 66}, 114009 (2002)
  [arXiv:hep-ph/0209030].

\bibitem{Tait:2000sh}
  T.~M.~P.~Tait and C.~P.~P.~Yuan,
  Phys.\ Rev.\  D {\bf 63}, 014018 (2000)
  [arXiv:hep-ph/0007298].

\bibitem{White:2009yt}
  C.~D.~White, S.~Frixione, E.~Laenen, F.~Maltoni,
  JHEP {\bf 0911}, 074 (2009).
  [arXiv:0908.0631 [hep-ph]].

\bibitem{Frixione:2008yi}
  S.~Frixione, E.~Laenen, P.~Motylinski, B.~R.~Webber, C.~D.~White,
  JHEP {\bf 0807}, 029 (2008).
  [arXiv:0805.3067 [hep-ph]].

\bibitem{Frixione:2005vw}
  S.~Frixione, E.~Laenen, P.~Motylinski, B.~R.~Webber,
  JHEP {\bf 0603}, 092 (2006).
  [hep-ph/0512250].

\bibitem{Chatrchyan:2011vp} 
  S.~Chatrchyan {\it et al.}  [CMS Collaboration],
  Phys.\ Rev.\ Lett.\  {\bf 107}, 091802 (2011)
  [arXiv:1106.3052 [hep-ex]].

\bibitem{Khachatryan:2014iya} 
  V.~Khachatryan {\it et al.}  [CMS Collaboration],
  JHEP {\bf 1406}, 090 (2014)
  [arXiv:1403.7366 [hep-ex]].

\bibitem{Aad:2014fwa} 
  G.~Aad {\it et al.}  [ATLAS Collaboration],
  Phys.\ Rev.\ D {\bf 90}, no. 11, 112006 (2014)
  [arXiv:1406.7844 [hep-ex]].

\bibitem{Aad:2012xca} 
  G.~Aad {\it et al.}  [ATLAS Collaboration],
  Phys.\ Lett.\ B {\bf 716}, 142 (2012)
  [arXiv:1205.5764 [hep-ex]].

\bibitem{Chatrchyan:2014tua} 
  S.~Chatrchyan {\it et al.}  [CMS Collaboration],
  Phys.\ Rev.\ Lett.\  {\bf 112}, no. 23, 231802 (2014)
  [arXiv:1401.2942 [hep-ex]].

\bibitem{eff}
A. Lucotte, A. Llères, D. Chevallier, ATLAS note ATL-PHYS-PUB-2007-005.

\bibitem{combined}
ATLAS and CMS Collaborations, ATLAS-CONF-2014-052, CMS-PAS-TOP-14-009

\bibitem{cms-pas-top-13-001}
CMS collab., 
CMS report CMS-PAS-TOP-13-001



\bibitem{Martinez:2007qf} 
  R.~Martinez, M.~A.~Perez and N.~Poveda,
  Eur.\ Phys.\ J.\ C {\bf 53}, 221 (2008)
  [hep-ph/0701098].

\bibitem{3-loop}
E.~P.~Shabalin, Yad. Eiz., {\bf 28}, 151 (1978); {\bf 31}, 1665 (1980); 
[Sov. J. Nucl. Phys., {\bf 28}, 75 (1978)]; I. B. Khriplovich, Phys. Lett. B {\bf 173},
193 (1986); Sov. J. Nucl. Phys. {\bf 44}, 659 (1986); YAFIA {\bf 44}, 1019 (1986);
A. Czarnecki and B. Krause, Phys. Rev. Lett. {\bf 78}, 4339 (1997).


\bibitem{mssm} 
  J.~M.~Yang and C.~S.~Li,
  Phys.\ Rev.\ D {\bf 54}, 4380 (1996)
  [hep-ph/9603442].

\bibitem{2hdm} 
  R.~Martinez and J.~A.~Rodriguez,
  Phys.\ Rev.\ D {\bf 65}, 057301 (2002)
  [hep-ph/0109109].
  
\bibitem{Gaitan:2015aia} 
  R.~Gaitan, E.~A.~Garces, J.~H.~M.~de Oca and R.~Martinez,
  arXiv:1505.04168 [hep-ph].

\bibitem{LH} 
  Q.~-H.~Cao, C.~-R.~Chen, F.~Larios and C.~-P.~Yuan,
  Phys.\ Rev.\ D {\bf 79}, 015004 (2009)
  [arXiv:0801.2998 [hep-ph]];
  L.~Ding and C.~-X.~Yue,
  Commun.\ Theor.\ Phys.\  {\bf 50}, 441 (2008)
  [arXiv:0801.1880 [hep-ph]].

\bibitem{unparticle} 
  R.~Martinez, M.~A.~Perez and O.~A.~Sampayo,
  Int.\ J.\ Mod.\ Phys.\ A {\bf 25}, 1061 (2010)
  [arXiv:0805.0371 [hep-ph]].




\bibitem{Rizzo:1995uv} 
  T.~G.~Rizzo,
  Phys.\ Rev.\ D {\bf 53}, 6218 (1996)
  [hep-ph/9506351].

\bibitem{Ayazi:2013cba} 
  S.~Y.~Ayazi, H.~Hesari and M.~M.~Najafabadi,
  Phys.\ Lett.\ B {\bf 727}, 199 (2013)
  [arXiv:1307.1846 [hep-ph]].

\bibitem{Fabbrichesi:2014wva} 
  M.~Fabbrichesi, M.~Pinamonti and A.~Tonero,
  Eur.\ Phys.\ J.\ C {\bf 74}, no. 12, 3193 (2014)
  [arXiv:1406.5393 [hep-ph]].

\bibitem{atwood}
  D.~Atwood, A.~Kagan and T.~G.~Rizzo,
  Phys.\ Rev.\ D {\bf 52}, 6264 (1995)
  [hep-ph/9407408].
\bibitem{haberl} 
  P.~Haberl, O.~Nachtmann and A.~Wilch,
  Phys.\ Rev.\ D {\bf 53}, 4875 (1996)
  [hep-ph/9505409].
\bibitem{hioki} 
  Z.~Hioki and K.~Ohkuma,
  Phys.\ Rev.\ D {\bf 83}, 114045 (2011)
  [arXiv:1104.1221 [hep-ph]];
  arXiv:1206.2413 [hep-ph].

\bibitem{saha} 
  D.~Choudhury and P.~Saha,
  Pramana {\bf 77}, 1079 (2011)
  [arXiv:0911.5016 [hep-ph]].
\bibitem{hioki2}
  Z.~Hioki and K.~Ohkuma,
  Eur.\ Phys.\ J.\ C {\bf 65}, 127 (2010)
  [arXiv:0910.3049 [hep-ph]];
  Eur.\ Phys.\ J.\ C {\bf 71}, 1535 (2011)
  [arXiv:1011.2655 [hep-ph]].
\bibitem{hesari} 
  H.~Hesari and M.~M.~Najafabadi,
  arXiv:1207.0339 [hep-ph].

\bibitem{gupta} 
  S.~K.~Gupta and G.~Valencia,
  Phys.\ Rev.\ D {\bf 81}, 034013 (2010)
  [arXiv:0912.0707 [hep-ph]];\\
  S.~K.~Gupta, A.~S.~Mete and G.~Valencia,
  Phys.\ Rev.\ D {\bf 80}, 034013 (2009)
  [arXiv:0905.1074 [hep-ph]].

\bibitem{Fabbrichesi:2013bca} 
  M.~Fabbrichesi, M.~Pinamonti and A.~Tonero,
  Phys.\ Rev.\ D {\bf 89}, no. 7, 074028 (2014)
  [arXiv:1307.5750 [hep-ph]].

\bibitem{Cheung:1995nt} 
  K.~-m.~Cheung,
  Phys.\ Rev.\ D {\bf 53}, 3604 (1996)
  [hep-ph/9511260].


\bibitem{Cheung:1996kc} 
  K.~-m.~Cheung,
  Phys.\ Rev.\ D {\bf 55}, 4430 (1997)
  [hep-ph/9610368].

\bibitem{Zhou:1998wz} 
  H.~-Y.~Zhou,
  Phys.\ Rev.\ D {\bf 58}, 114002 (1998)
  [hep-ph/9805358].

\bibitem{Englert:2012by}
  C.~Englert, A.~Freitas, M.~Spira and P.~M.~Zerwas,
  arXiv:1210.2570 [hep-ph].

\bibitem{Hewett:1993em} 
  J.~L.~Hewett and T.~G.~Rizzo,
  Phys.\ Rev.\ D {\bf 49}, 319 (1994)
  [hep-ph/9305223].

\bibitem{b2sg} 
  R.~Martinez and J.~A.~Rodriguez,
  Phys.\ Rev.\ D {\bf 55}, 3212 (1997)
  [hep-ph/9612438];\\
  J.~F.~Kamenik, M.~Papucci and A.~Weiler,
  Phys.\ Rev.\ D {\bf 85}, 071501 (2012)
  [arXiv:1107.3143 [hep-ph]].

\bibitem{Godbole:2006tq} 
  R.~M.~Godbole, S.~D.~Rindani and R.~K.~Singh,
  JHEP {\bf 0612}, 021 (2006)
  [hep-ph/0605100];

  \bibitem{Grzadkowski:1999iq} 
  B.~Grzadkowski and Z.~Hioki,
  Phys.\ Lett.\ B {\bf 476}, 87 (2000)
  [hep-ph/9911505];
  S.~D.~Rindani,
  Pramana {\bf 54}, 791 (2000)
  [hep-ph/0002006];
    B.~Grzadkowski and Z.~Hioki,
  Phys.\ Lett.\ B {\bf 529}, 82 (2002)
  [hep-ph/0112361];
  Phys.\ Lett.\ B {\bf 557}, 55 (2003)
  [hep-ph/0208079];\\
  Z.~Hioki,
  hep-ph/0210224.

\bibitem{Vermaseren:2000nd}
  J.~A.~M.~Vermaseren,
  arXiv:math-ph/0010025.

\bibitem{cteq6}
  J.~Pumplin, A.~Belyaev, J.~Huston, D.~Stump and W.~K.~Tung,
  JHEP {\bf 0602}, 032 (2006)
  [arXiv:hep-ph/0512167].

\end{thebibliography}
\end{document}